\documentclass[twocolumn]{svjour3}          
\pdfoutput=1
\smartqed  

\usepackage{graphicx}
\usepackage{marvosym}
\usepackage{mathptmx}
\usepackage{siunitx}
\usepackage{amsmath,amssymb,multirow}
\usepackage{color}
\usepackage{colortbl}            
\usepackage[sort]{natbib}
\bibpunct{(}{)}{,}{a}{}{;}

\journalname{Submitted}

\begin{document}

\title{Jet mixing optimization using machine learning control}


	
	\author{Zhi Wu$^1$ \and
                Fan Dewei$^1$ \and
                Yu Zhou$^1$ \and
                Ruiying Li$^{2}$ \and
		Bernd R. Noack$^{3,1,4}$}
	
	\institute{
\Letter~Yu Zhou\\
\email{yuzhou@hit.edu.cn}\\
\Letter~Bernd R. Noack\\
\email{Bernd.Noack@limsi.fr}\\ \\
$^1$ Institute for Turbulence-Noise-Vibration Interaction and Control,
Harbin Institute of Technology, Shenzhen Campus, Shenzhen,
People’s Republic of China\\ \\
$^2$ Institut PPRIME, CNRS -- Universit\'e de Poitiers -- ISAE-ENSMA,
86962 Futuroscope Chasseneuil, France\\ \\
$^3$ LIMSI-CNRS,  Rue John von Neumann, Campus Universitaire d'Orsay, B\^at 508,
F-91403 Orsay, France\\ \\
$^4$ Institut f\"ur Str\"omungsmechanik und Technische Akustik (ISTA),
Technische Universit\"at Berlin,
Stra{\ss}e des 17. Juni 134, D-10623 Berlin, Germany}

\date{Received: date / Accepted: date}

\maketitle

\begin{abstract}
We experimentally optimize mixing
of a turbulent round jet using machine learning control (MLC) following \citet{Li2017ef}.
The jet is manipulated with one unsteady minijet blowing in wall-normal direction close to the nozzle exit.
The flow is monitored with two hotwire sensors.
The first sensor is positioned on the centerline 5 jet diameters downstream of the nozzle exit,
i.e.\  the end of the potential core,
while the second is located 3 jet diameters downstream and displaced towards the shear-layer.
The mixing performance is monitored with mean velocity at the first sensor.
A reduction of this velocity correlates with increased entrainment near the potential core.
Machine Learning Control (MLC) is employed to optimize sensor feedback,
a general open-loop  broadband  frequency actuation
and combinations of both.
MLC has identified the optimal periodic forcing with small duty cycle as the best control policy
employing only 400 actuation measurements, each lasting for 5 seconds.
This learning rate is comparable if not faster than
typical optimization of periodic forcing with two free parameters (frequency and duty cycle).
In addition, MLC results indicate that neither new frequencies
nor sensor feedback improves mixing further---contrary to many of other turbulence control experiments.
The optimality of pure periodic actuation may be attributed
to the simple jet flapping mechanism in the minijet plane.
The performance of
sensor feedback is shown to face a challenge for small duty cycles.
The jet mixing results demonstrate the untapped potential of MLC
in quickly learning optimal general control policies,
even deciding between open- and closed-loop control.

\end{abstract}

\setcounter{tocdepth}{3}

\section{Introduction}
\label{Sec:Introduction}
The enhancement of jet mixing is important to many industrial applications.
One example is the production of polymers
where additional substances are added in the main stream by jets
and have to be rapidly mixed downstream.
Similar mixing examples can be found in food industry.
Another application is combustion in aeroengines:
The first row of dilution jets enhances combustion by mixing
while the second downstream row  cools the fluid
to prevent thermal damage to the turbine.
Staying with aeroengines, the targeted homogeneous mixing
of injected fuel in the airstream is affected by two-phase jet mixing.
Carrier airplanes reply on enhanced jet mixing with flapping
to prevent the burning of their tail after landing.

Consequently, jet mixing control has drawn significant attention in the past few decades.
Jet mixing control can be classified into passive control and active control
based on whether additional energy input is needed.
For example, the use of non-circular nozzles \citep{Gutmark1999arfm}
or the deployment of tabs at the nozzle exit \citep{Bradbury1975jfm,Zaman1994pf} are passive control techniques.
These devices can display  impressive performance for the design conditions.
Yet, the techniques are typically permanent fixtures
which are not readily modified or removed.
In addition, the control performance may deteriorate departing from the design condition.
In contrast, the active control
may potentially achieve higher performance for a large range of operating conditions.
Examples are acoustic excitation \citep{Zaman1980jfm}, plasma actuators \citep{Samimy2007jfm}, synthetic jet \citep{Ho1987jfm},
oscillating boundaries based on piezo-electric actuators \citep{Wiltse1993jfm},
steady and unsteady control jets \citep{Davis1982aiaaj,Yang2016jfm,Zhou2012aiaaj}.
From an industrial perspective,
actuators and sensors become increasingly more reliable and cheaper,
i.e.\ more attractive for applications.
Thus, active flow control enjoys increasing progress from many fronts,
like hardware development, control logic and modeling.

Active control techniques can be performed in an open-loop or closed-loop manner.
By definition, closing the loop with sensors increases the opportunity space of actuation
and---properly set up---should improve performance.
The loop may be closed for in-time response to coherent structures
or for adapting in response to changing flow conditions.
Note that, `in-time' means that the actuation responds on a time-scale much smaller
than that of the physical process, while `adapting' means that the change of the actuation parameter is slow as compared
to the physical process time-scale \citep{Brunton2015amr}.
Many closed-loop control schemes have been proposed and investigated,
 as discussed in references \citet{Brunton2015amr,Choi2008arfm,Collis2004pas}.
Closed-loop control may be classified into model-based or model-free approaches,
depending on whether the law is derived from a plant model or only based on the plant response.
Many computational flow control studies are based on local linearization of a Navier-Stokes based model.
Linear models may be also identified in black-box manner from input-output data sequences \citep{Rapoport2003jpp}
or via a reduced-order model of the fluid dynamics \citep{Choi2008arfm}.
The challenge to this approach is the nonlinear dynamics of turbulence
displaying myriad of frequency-crosstalk mechanisms,
like the change of the base flow by coherent structures
or the turbulence cascade to smaller and smaller vortices
with higher and higher frequencies.
These important frequency crosstalk mechanisms are ignored in linear control strategies.
They are also difficult to encapsulate in low-dimensional control-oriented models.
This challenge motivates model-free control techniques which do not rely on a dynamical model of the fluid system.
The majority of the experimental turbulence control studies
rely on adaptive variation of one or few actuation parameters,
like the amplitude or frequency of suction or blowing.
Examples include physics-based methods, like \citet{Pastoor2008jfm,Wu2016ef,Zhang2004pf},
extremum and slope-seeking control method \citep{Wu2016ef,Becker2007aiaaj,Brackston2016ef,Maury2012ija,Wu2015ef}
and Machine Learning Control (MLC) \citep{Brunton2015amr,Duriez2016book,Li2017ef} allowing for a very rich set of possible control laws.
All approaches have been widely applied in turbulence experiments.
For example, in the jet mixing enhancement with one pulsed minijet,
extremum-seeking control (ESC) has been applied to obtain automatically
and rapidly the optimal excitation frequency $f_{e,opt}$ of the minijet
as monitored by the maximum decay rate $K_{max}$ of the jet centerline mean velocity.
This ESC adapts a working open-loop control \citep{Wu2016ef,Fan2017aiaaj}.
\cite{Zhang2004ejmb} used an in-time proportional-integral-derivative (PID) control
to suppress the flow-induced vibration on a square cylinder by generating a very small cylinder surface oscillation,
achieving a control performance which outperforms optimized open-loop control.

Hitherto, in-time turbulence control is a largely unexplored Terra Incognita.
This is particularly true for jet mixing
displaying a rich set of temporal and spatial scales.
In this study, we seize this opportunity
exploring a very rich set of control laws for jet mixing optimization---incorporating all major open- and closed-loop actuations.
The recently discovered and enabling MLC framework has not yet been applied to this configuration.

In the current work,
we optimize jet mixing with a single minijet actuator and two downstream hotwire sensors.
As cost function, 
the averaged centerline velocity after the potential core is taken.
For the control logic,
we employ MLC based on linear genetic programing
incorporating sensor-based feedback and multi-frequency forcing as well as combinations thereof.
We follow a methodologically similar MLC study for drag reduction of an Ahmed body \citep{Li2017ef}.
MLC significantly increases the range of possible control laws as compared to ESC.
Moreover MLC does not rely on the qualitative knowledge of steady-state maps
and could find the global optimal value even when multiple extrema exist.

The manuscript is structured as follows.
Section~\ref{Sec:Setup} describes the experimental setup.
The control logic and associated MLC algorithm is outline in Sect.~\ref{Sec:MLC}.
In Sect.~\ref{Sec:ClosedLoopControl} and Sect.~\ref{Sec:OpenLoopForcing},
we detail the experimental control with sensor-based feedback and multi-frequency forcing, respectively.
In Sect.~\ref{Sec:Discussion} the mixing performance of different control schemes is physically explained.
The conclusions Sect.~\ref{Sec:Conclusions} summarize the MLC study and preview future developments.

\section{Experimental set-up}
\label{Sec:Setup}
\subsection{Jet facility and actuator system}
\label{jet facilities}
 \begin{figure*}
 	\centering
 	\includegraphics[width=0.80\textwidth,keepaspectratio]{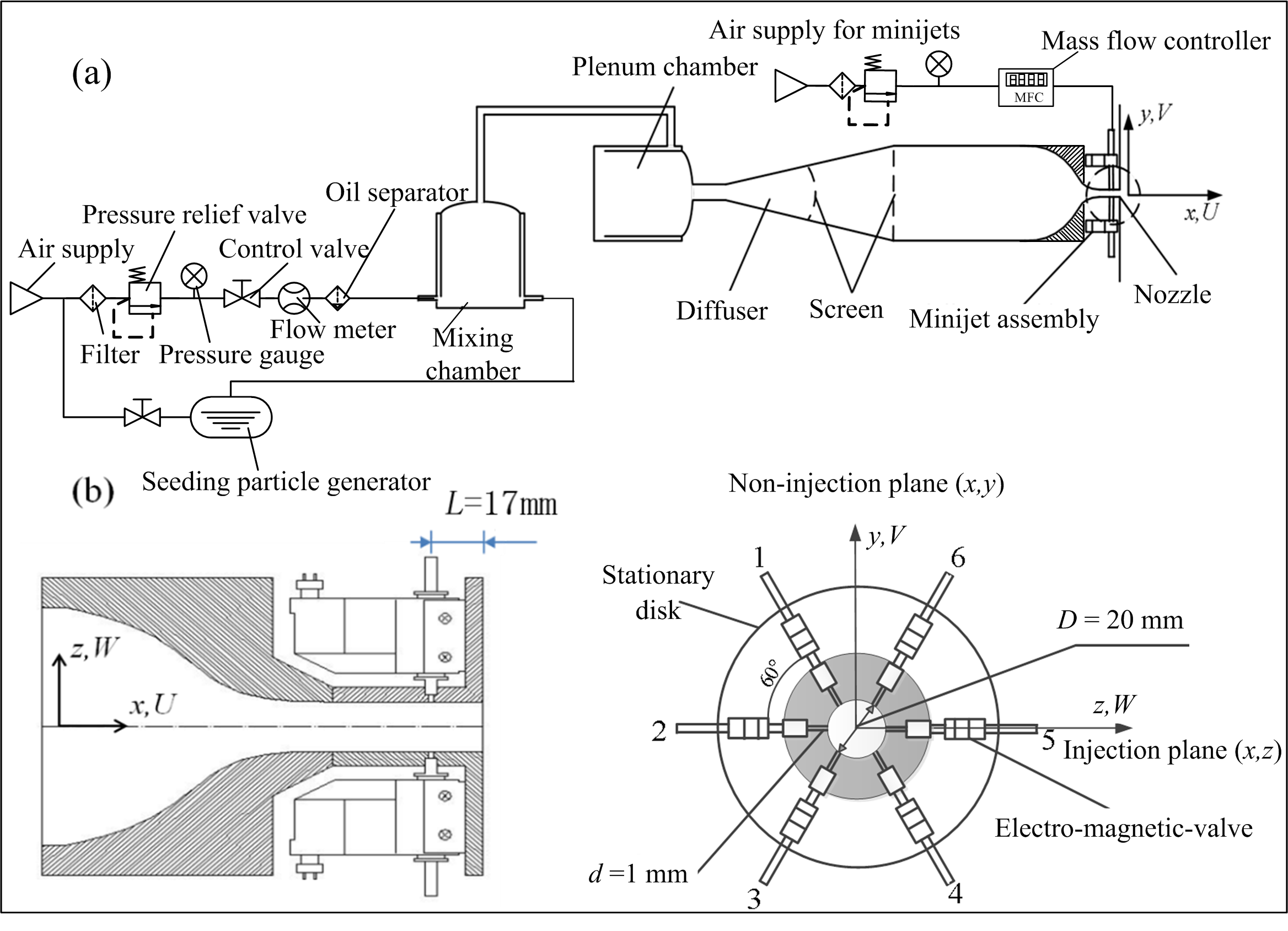}
 	\caption{Schematic of the experimental setup: (a) main-jet assembly; (b) minijet assembly.}
 	\label{Fig:Setup}
 \end{figure*}
The jet control platform,
consisting of an axisymmetric main jet and a minijet assembly,
is the same as that used in \cite{Fan2017aiaaj}.
Figure  \ref{Fig:Setup}a shows the schematic diagram of the air jet facility.
The air for both main jet and minijet comes from the same compressed air supply with a 12 bar gauge pressure.
The compressed air firstly passes through a mixing chamber in the case of particle image velocimetry (PIV)
or flow visualization measurements. After that, it passes through a tube, a plenum chamber,
a 300 mm--long diffuser of $15^{\circ}$ in half angle, two fine screens (7 mesh/cm)
and a cylindrical settling chamber of 400 mm in length and 114 mm in the inner diameter. 
The nozzle contraction contour follows equation $R = 57-47\sin1.5(90-9x/8)$ (mm),
as used in \cite{Zhou2012aiaaj} and \cite{Wu2016ef}.
The nozzle was extended with a 47 mm-long smooth tube of diameter $D=20$ mm.
The Reynolds number $Re_D=U_j D/\nu$ of the main jet is fixed at $8000$,
where $U_j$ is the jet centreline velocity measured at the exit of the nozzle extension and $\nu$ is the kinematic viscosity of air.
Figure \ref{Fig:Setup}b shows the schematic diagram of the pulsed minijet assembly.
There are six orifices of 1 mm in diameter drilled radially for the minijets,
17 mm upstream of the jet exit.
To minimize the resistance,
the orifices are suddenly expanded to a diameter of 4 mm before being connected via a short plastic tube to electromagnetic-valves (Koganei K2-100SF-09-LL),
which are used to produce a pulsed minijet (Fig. \ref{Fig:Setup}b).
The electromagnetic-valves can generate the pulsed jet in ON/OFF mode within the frequency range $[0, 500]$ Hz.
In the present study, we control only the ON/OFF of the electromagnetic-valves.
The maximum frequency of the minijet is $500$ Hz,
exceeding $3f_0$ at $Re_D = 8000$,
where $f_0=135$ Hz is the dominant frequency of the uncontrolled jet.
Only one single minijet injection was investigated presently.
The mass flow rate of the minijet varied via a mass flow controller (FLOWMETHOD FL-802)
with a range of 7 Standard Liter Per Minute (SLPM),
whose experimental uncertainty is no more than 1\%.
The mass flow rate of the minijet is fixed at 1.3 l/min for all the current investigations, corresponding to $C_m$=1.2\%, where the $C_m$ is the mass flow rate ratio of the main jet to the minijet.

\subsection{Velocity measurement and flow visualization facilities}
\label{measurement facilities}
The origin of the coordinate system is defined at the centre of the jet exit,
with the $x$- and $z$-axes along the streamwise and the radial minijet directions, respectively, and the $y$ axis is normal to the $(x, z)$ plane,
following the right-hand rule.
The $(x, z)$ and $(x, y)$ planes are hereinafter referred to as the injection and non-injection planes, respectively.
In this paper, an asterisk superscript denotes normalization by $D$ or/and $U_j$.

Two tungsten wires of $5 \> \mu$m in diameter,
operated on a constant temperature circuit (Dantec Streamline) at an overheat ratio of 1.8,
is placed at $(x^*,y^*) = (3, 0.25)$ and $(5, 0)$ to measure the streamwise velocities $u_{3D}$ and $u_{5D}$, respectively.
Note that the slight deviation from centreline for hotwire $1$, which is for measuring the $u_{3D}$,
is used to prevent its influence on hotwire $2$.
The output signal of the hotwire anemometer is offset, amplified, and filtered at a cut-off frequency of 500 Hz
before being digitized and saved in a PC by a \textit{National Instrument} multifunction I/O Device (PXIe-6356).
The sampling frequency is 1 kHz for all  experiments.
The hotwires were calibrated at the jet exit using a pitot tube and a micromanometer (Furness Controls FCO510).
The experimental uncertainty of the hotwire measurement is estimated to be less than 2\%.

A planar high-speed PIV system, including a high speed camera (LaVision ImagerproHS4M, $2016 \times 2016$ pixels resolution)
and pulsed laser source (Litron LDY304-PIV, Nd:YLF, 120 mJ/pulse) is deployed for flow visualization in the $(x, z)$ and $(x, y)$ planes.
An oil droplet generator (TSI MCM-30) is used to generate fog for seeding flow.
The seeding particles are supplied into the mixing chamber (Fig. \ref{Fig:Setup}a) to mix with air.
Flow illumination is provided by a laser sheet of 1 mm in thickness generated by a pulsed laser source of 120 mJ via a cylindrical lens.
Particle images are captured at a sampling rate of 250 Hz.

\subsection{Real-time system}
\label{RT system}
The real-time control is realized by a \textit{National Instrument} PXIe-6356 multifunction I/O Device
running at a sampling rate of $F_{RT}$=1 kHz, where a LabVIEW Real-Time module is used to process the program.
Sensor data acquisition and control command generation for open- and closed-loop control
are performed at the same sampling rate.
For the effective working of the actuator,
a verification is performed before sending the command to the actuators to ensure that the ON/OFF command lasts at least 1 ms.

The available periodic frequencies $f$ consistent with $F_{RT}$
can be derived from $f=F_{RT}/N_{sp}$,
where $N_{sp}$ is the number of sampling points in one time period  $1/f$.
The working frequency range of actuators ($[0,500]$ Hz)
imposes a minimum value for $N_{sp}$,  being $N_{sp}\geq 2$.
For a given $f$, the possible duty cycle $DC$ can be deduced from $DC=m/N_{sp}$, $m=1,\ldots,N_{sp}-1$.
The value of $m$ starts from 1 and ends at $N_{sp}-1$ to ensure a response time of 1 ms for an effective working of the actuators.
Thus, the number of possible duty cycles $N_{DC}$ for a given $f$ is $N_{DC}=N_{sp}-1=F_{RT}/f-1$,
which increases with $N_{sp}$ and decreases with $f$.
This process is similar to that used in \cite{Li2017ef}.

\begin{figure}
\begin{center}
 	\includegraphics[width=0.45\textwidth,keepaspectratio]{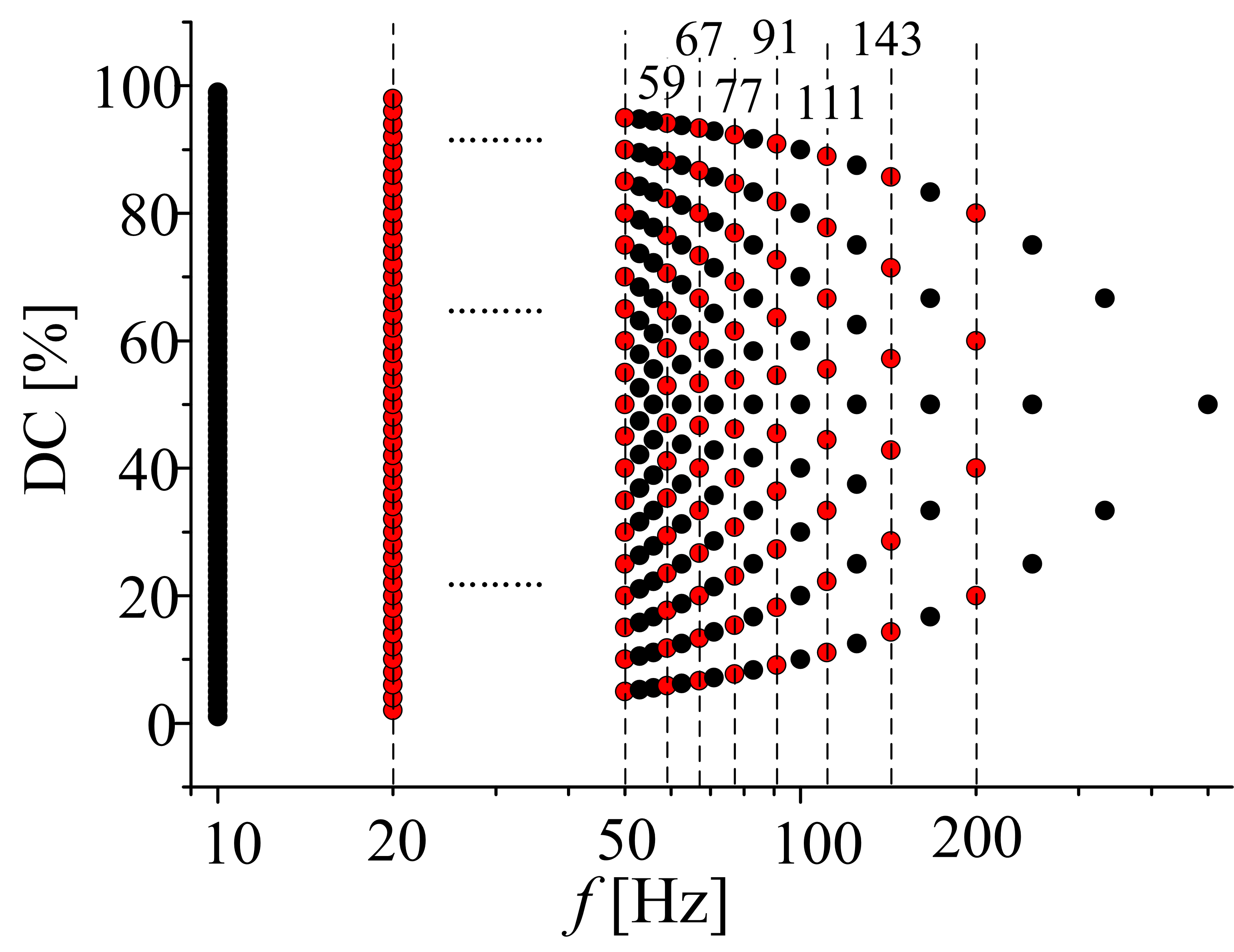}
\end{center}
\caption{Frequency selection for the multi-frequency forcing ansatz \eqref{Eqn:MultiFrequencyForcing}.
Each circle corresponds to a frequency and duty cycle achievable with a $F_{RT}=1$ kHz data acquisition system.
The horizontal lines and numbers indicate the selected frequencies for harmonic inputs.
 Red dots highlight the cases considered in this study.
 Black dots represent other combinations of $f$ and $DC$ consistent with $F_{RT}$.
Combinations between [10 20]Hz and [20 50]Hz are not explicated for reasons of figure clarity.
See Figure 3 of \cite{Li2017ef} for a comparison with similar selection process.}
\label{Fig:Frequencies}
\end{figure}

Figure \ref{Fig:Frequencies} displays the permitted  frequencies and duty cycles and shows the manually selected frequencies which
allow for a locally maximum number of duty cycles.
The number of possible duty cycles decreases as the frequency increases due to the limited sampling points in one period.
Note that Fig.~\ref{Fig:Frequencies} includes all the possible $DC$ and $f$ within the range $[10,500]$.
The red filled circles highlight the selected periodic forcing cases considered in the following, which contains the most interesting frequency range $[20,200]$ Hz.

\section{Machine learning control (MLC)}
\label{Sec:MLC}

The mixing enhancement of a turbulent jet
involves a large range of temporal and spatial scales
with complex nonlinear interactions.
Model-based control as used for the stabilization of steady laminar flows
faces fundamental challenges for such turbulent flow
and has, to the best of the authors knowledge,
not been presented for experimental mixing enhancement of turbulence.
Instead, we follow Dracopoulos \& Kent's  pioneering work \citep{Dracopoulos1997nca},
in which control design is framed as regression problem
and solved with one of the most powerful method
of machine learning: genetic programming.
This strategy, referred to as \emph{Machine Learning Control (MLC)}
in recent literature \citep{Brunton2015amr,Duriez2016book}
 has been applied with large success for a range
of turbulence control experiments
 \citep{Noack2017fssic}.
MLC has continually outperformed existing control strategies
often exploiting surprising nonlinear frequency cross-talk mechanisms.

In this section,
the employed MLC implementation is described.
First (Sect.~\ref{Sec:CostFunction}),
the control problem is framed as optimization of a cost function.
Then (Sect.~\ref{Sec:ControlLaw}), a general control ansatz is proposed.
This ansatz comprises nonlinear sensor-feedback with signal history,
multi-frequency forcing and combinations thereof.
Now, the search for an optimal control law
can be formulated as regression problem (Sect.~\ref{Sec:RegressionProblem}).
Finally (Sect.~\ref{Sec:GeneticProgramming}), the employed
linear genetic programming is detailed as powerful regression solver,
including the parameters of this evolutionary algorithm.

\subsection{Cost function}
\label{Sec:CostFunction}
Good jet mixing is associated with large entrainment of the ambient flow
into the high-momentum jet fluid emanating from the orifice.
This entrainment reduces the streamwise velocity on the centerline.
Following earlier work \citep{Wu2016ef},
we take the streamwise velocity $u_{5D}$
five jet diameters downstream
as mixing indicator.
This location is approximately at the end of the potential core.
The smaller this velocity,
the better is the mixing between the high-momentum jet
and the surrounding ambient fluid.
In the cost function,
the centerline velocity is normalized
with the maximum jet velocity at the orifice
\begin{equation}
\label{Eqn:CostFunction}
J =  \frac{\overline{u_{5D}}}{U_j}.
\end{equation}
Here, the overbar denotes a time average.
This non- dimensionalization allows to compare jet mixing
for a range of operating conditions.

The minimization of the cost function corresponds
to the maximization of the jet centerline decay rate,
\begin{equation}
\label{Eqn:DecayRate}
\overline{K} = \frac{U_j - \overline{u_{5D}}}{U_j} = 1 - J.
\end{equation}
The decay rate quantifies the streamwise velocity deficit  on the centerline. $\overline{K}$ is correlated approximately with an equivalent jet width $R_{eq}=\left[R_HR_V\right]^{1/2}$, where $R_H$ and $R_V$ are the jet half-widths in two orthogonal planes \citep{Zhou2012aiaaj}, that is, $\overline{K}$ is directly correlated to the entrainment rate.

\subsection{Ansatz for control law}
\label{Sec:ControlLaw}
We search to optimize actuation
in a very general ansatz for the control law
comprising, for instance,
multi-frequency forcing
and sensor-based feedback with signal history.
In the following the actuation command is denoted by $b$.
It can take binary values $1$ and $0$ depending if minijet is on or off.
The minijet velocity scales approximately
with the inverse of the the duty cycle.

\subsubsection{Multi-frequency forcing}
In similar experiments \citep{Fan2017aiaaj},
a periodic forcing with a frequency $\omega^{\star}=2 \pi f^{\star}$, $f^{\star} \approx 67$ Hz
was found to be very effective.
The corresponding open-loop control law reads
\begin{equation}
\label{Eqn:PeriodicForcing}
b (t) = H \left( \sin ( \omega^{\star} t )- k \right),
\end{equation}
where $H$ represents the Heaviside function and $k$ controls the duty cycle.
The larger $k \in (-1,1)$,
the smaller the duty cycle.
Following \cite{Li2017ef},
a much more general multi-frequency forcing is considered,
generated here by 9 harmonic functions
$h_i = \sin  ( 2 \pi f_i t ) $, $i=1,\ldots,9$.
The frequencies are selected based on the data acquisition frequency of $1$ kHz.
With finite data acquisition frequency,
only discrete frequencies
with discrete duty cycles are possible as displayed in Fig.~\ref{Fig:Frequencies}.

We comprise the harmonic functions into a vector-valued frequency generator
\begin{equation}
\label{Eqn:HarmonicFunctionVector}
\vec{h}(t) = \begin{bmatrix} h_1 & h_2 & \ldots & h_9 \end{bmatrix}^\dagger
\end{equation}
where the  $\dagger$ superscript denotes the transpose,
$h_i = \cos  ( 2 \pi f_i t ) $, and
$f_i =  {20, 50, 59, 67, 77, 91, 111, 143, 200}$ Hz for $i=1,\ldots.9$ respectively.
We do not include sinusoidal functions at the same frequencies.
In the open-loop actuation literature,
 phase differences are only found to be important for few frequency ratios,
e.g.\ harmonic and subharmonic components of the mixing layer \citep{Monkewitz1988jfm}.
Most Lissajous figures with $h_i, h_j$ densely fill out the square $[-1,1] \times [-1,1]$,
indicating the phase difference cannot be expected to have an effect.
The open-loop multiple forcing actuation is performed with
\begin{equation}
\label{Eqn:MultiFrequencyForcing}
b (t) = B ( \vec{h} (t) ).
\end{equation}
If $B$ were linear function --- putting the binary nature of actuation aside ---
the resulting actuation command can exhibit the input frequencies.
If $B$ were a pure quadratic function, all different frequencies may appear.
In case of a general nonlinear function, a large range of frequencies can be generated.
For instance, $h_1^{10} -1/2$ generates 
a harmonic function with frequency $10 f_1$.
Hence, the main frequency limitation of the general ansatz \eqref{Eqn:MultiFrequencyForcing}
is not caused by the ansatz but by the actuator performance.

\subsubsection{Sensor-based feedback}
The considered feedback is based on the hot-wire signal $u_{5D}$
used for the cost function \eqref{Eqn:CostFunction}
and on the slightly displaced hot-wire measurement three diameters downstream $u_{3D}$
monitoring the shear-layer vortices.
The feedback signals are based on the
Reynolds decomposition
into a one period average denoted by $\langle \cdot \rangle_T$
and a fluctuation, $u_{5D}= \langle u_{5D} \rangle_T + u_{5D}^{\prime}$
and $u_{3D}= \langle u_{3D} \rangle_T + u_{3D}^{\prime}$.
The period $T$ is taken from the best periodic forcing frequency $67$ Hz.
In addition, the signals are normalized with the jet velocity at the nozzle exit
to arrive at more robust control laws
for a range of operating conditions, e.g.\ different jet velocities.
The resulting feedback argument reads
\begin{equation}
\label{Eqn:Sensor}
\vec{s}(t) =    \begin{bmatrix} s_1(t) \\ s_2(t) \\ s_3(t) \\ s_4(t)  \end{bmatrix}
= \frac{1}{U_j} \begin{bmatrix} u_{3D} (t) \\
                                u_{5D} (t) \\
                                         u_{3D}^{\prime}(t) \\
                                         u_{5D}^{\prime}(t) \\
                \end{bmatrix}.
\end{equation}
Note that $J = \overline{s_2}$.

The ansatz for the sensor-based feedback law has the form:
\begin{equation}
\label{Eqn:SensorFeedback}
b(t) = B \left( \vec{s} (t) \right).
\end{equation}

\subsubsection{Generalized feedback control}

A natural generalization of the previous approaches reads
\begin{equation}
\label{Eqn:GeneralizedFeedback}
b (t) = B ( \vec{s} (t), \vec{h}(t) ).
\end{equation}
The sensor-based feedback 
might, for instance,
 control the duty cycle of an actuation frequency.
In Sect. \ref{Sec:ClosedLoopControl} and \ref{Sec:OpenLoopForcing},
we will explore jet mixing performance for a variety of control laws.

\subsection{Control design as model-free regression problem}
\label{Sec:RegressionProblem}

The cost function $J$ evidently depends
on the chosen control logic $B \left( \vec{s}, \vec{h} \right) $.
We search for a control law which minimizes $J$,
\begin{equation}
\label{Eqn:RegressionProblem}
B^{\star} = \underset{B}{\hbox{\rm arg min}} \> J
  \left[ B \left (\vec{s}, \vec{h} \right ) \right].
\end{equation}

It is important to realize that this optimization task
is a \emph{regression problem of the second kind}:
the optimal actuation command  $b$
for given sensor input  $\left(\vec{s}, \vec{h} \right)$ is not known
and hence can not be learned directly,
as in a \emph{regression problem of the first kind}.
One can only judge the performance of a control law based on the cost function.
This appears to be a very subtle distinction
but practically excludes many known regression solvers of machine learning.
For instance artificial neural networks
and deep learning typically
require `guidance' for the optimal mapping output.
Such actuation may be available from an optimal control
using the full Navier-Stokes equations \citep{Lee1997pf}.
For experiments, however, we rely on the control performance alone.

\subsection{Linear genetic programming as regression solver}
\label{Sec:GeneticProgramming}

Following \cite{Li2017ef},
the control optimization problem \eqref{Eqn:RegressionProblem}
is solved using \emph{linear genetic programming (LGP)}.
LGP is a powerful regression technique of machine learning
which can optimize general nonlinear mappings, like the control law.
We refer to the exquisite textbook of Wahde \citep{Wahde2008book}
for a quick overview on a spectrum of evolutionary algorithms including LGP,
to the detailed LGP textbook by Brameier \& Banzhaf \citep{Brameier2007book}
and to the first turbulence control applications \citep{Li2017ef}.
In this section, LGP is sketched and the key parameters are listed.
\begin{figure}
\begin{center}
\includegraphics[width=0.45\textwidth]{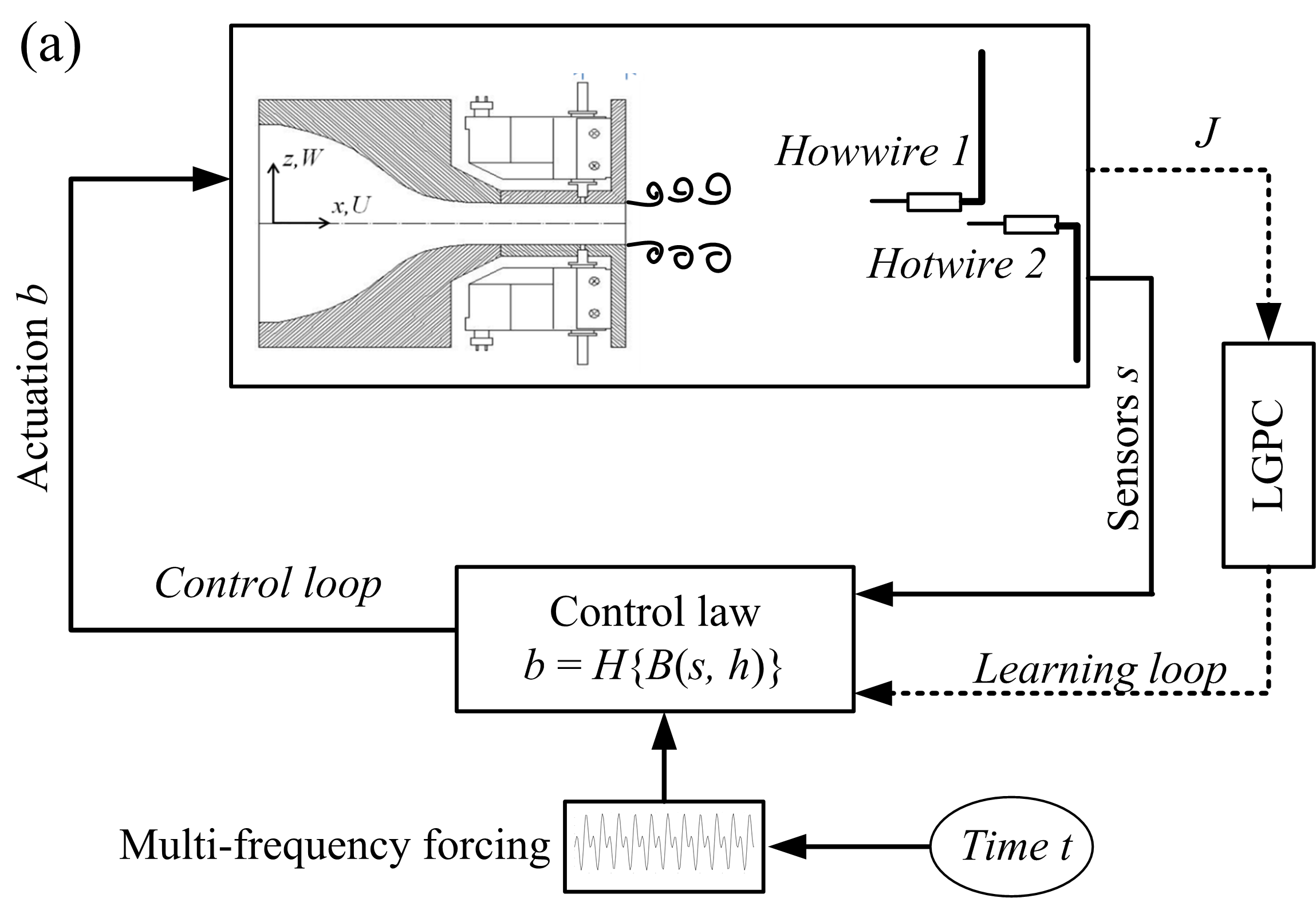} \\
\includegraphics[width=0.45\textwidth]{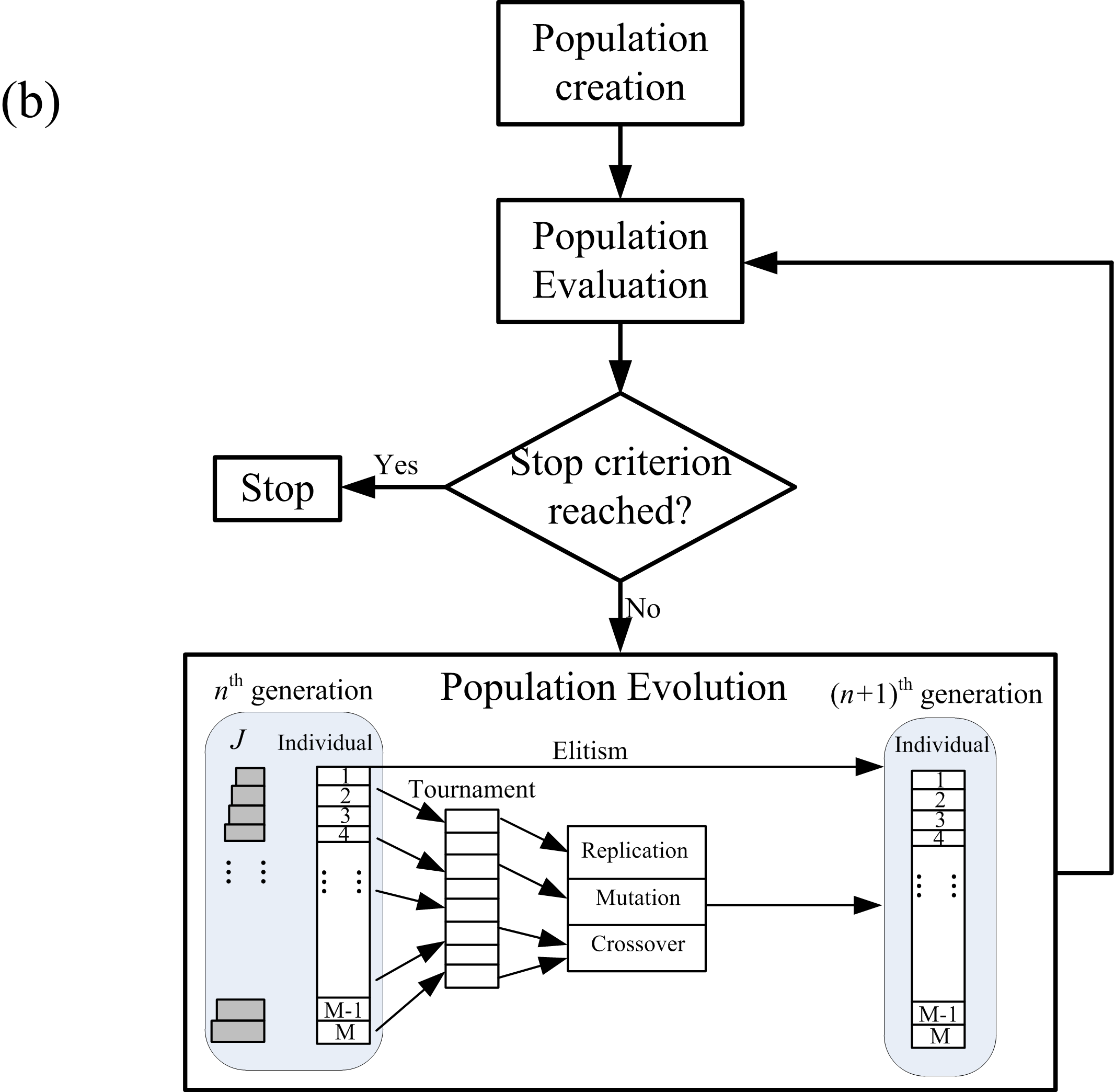} \\
\end{center}
\caption{Principle sketch of machine learning control.
(a) shows the fast inner control loop with the jet mixing plant and
the slow outer evaluation / learning loop.
(b) outlines the evolutionary algorithm behind the learning of the best control law.}
\label{Fig:MLC}
\end{figure}

LGP control starts like a Monte-Carlo approach
with $I$ random control laws, referred to as \emph{individuals}:
\begin{equation}
\label{Eqn:FirstGeneration}
 b = B_i^{(1)}, \quad i=1, \ldots, I.
\end{equation}
The superscript (1) represents the first generation. The jet mixing experiment
grades the performance of each law $J_î^{(1)}$
in the plant.
Figure \ref{Fig:MLC} shows the plant, the fast inner control loop
and the slow outer evaluation / learning loop.
Then, the individuals are re-numbered and ranked in order of performance,
\begin{equation}
\label{Eqn:CostFunctionOrder}
J_1^{(1)} \le J_2^{(1)} \le \ldots \le J_{I}^{(1)}.
\end{equation}
The best $N_e$ individuals are adopted in the new generation,
\begin{equation}
B_i^{(2)} = B_i^{(1)}, \quad i=1,\ldots,N_e.
\end{equation}
This operation is called \emph{elitism}.
The remaining $I-N_e$ individuals of the new generation
are determined with a random sequence
of three genetic operations:
\emph{crossover},
\emph{mutation}, and
\emph{replication}
with probabilities $P_c$, $P_m$ and $P_r$, respectively.
Crossover has two arguments (individuals) and breeds two new individuals
exchanging equally sized part of `genes' from both individuals.
This operation tends to yield better individuals,
e.g.\  \emph{exploit populated local minima}.
Mutation has a single argument and replaces part of the `genes' rather randomly. 
This operation might  \emph{explore new local minima}.
Replication has one argument which is copied unaltered into the new generation.
Evidently, this operation has a memory effect.
The argument of these operations is decided in a tournament.
$N_t$ individuals of the graded generations are chosen with equal probability for the tournament.
The genetic operation takes the best or the best two individuals,
i.e.\ there is a bias towards processing better individuals
but low performing individuals are not completely ignored.

These iterations are performed until convergence of the best performing individuals
or the end of the measurement time is reached (see Fig.\ \ref{Fig:MLC}b).
Let $N_g$ be the number of the evaluated generations before termination.
The best control law of the last generation $B_1^{(N_g)}$
is taken as the solution of the regression problem \eqref{Eqn:RegressionProblem}.

In LGP,
each individual consists of a set of instructions
using elementary operations on a register $r_j$, $j=1,\ldots,N_s+N_b+N_c$,
with the first $N_s$ values for the input signals,
the next $N_b=1$ values for the actuation commands,
and $N_c$ constants.
For sensor-based feedback \eqref{Eqn:SensorFeedback} there are four input signals $N_s=4$,
for multi-frequency forcing \eqref{Eqn:MultiFrequencyForcing} $N_s=9$
and for the generalized feedback \eqref{Eqn:GeneralizedFeedback} $N_s=4+9=13$.
Before any operation,
the input registers are initialized with the argument values
and the actuation register is zeroed.
The $k$th instruction is coded as integer matrix $B_{kl}$, $l=1,\ldots,4$.
Here, $B_{k3}$ denotes the index of the operation, e.g. `$1$' for `$+$',
$B_{k1}$, $B_{k2}$ represents the index of the input registers
and $B_{k4}$ the index of the output registers (excluding the constants).
Thus, $B_{k1}=2$, $B_{k2}=3$, $B_{k3}=1$ and $B_{k4}=5$ corresponds to $r_5=r_1+r_2$.

The employed parameters of LGP are listed
in Table \ref{tab:LGP_param}.
The same or very similar parameters have been chosen in dozens of other turbulence control experiments,
flow control simulations or dynamical systems control \citep{Duriez2016book}.
The performance of the resulting machine learning control
was not critically dependent on any of these parameters.
\begin{table}
	\centering	
	\begin{tabular}{lr}
		\hline
Parameters  & Value  \\	\hline
Population size & $I=100$ \\
Elitism & $N_e =1$ \\
Crossover & $P_c =70$ \% \\
Mutation & $P_c =20$ \% \\
Replication & $P_c =10$ \% \\
Operations & $+, -,  \times, \div,  x^2 $\\
           & $\sin, \tanh, \log, \exp $\\
Constants & $N_c=6$ \\
Constant range & $[-1,1]$\\ \hline	
	\end{tabular}
	\caption{LGPC parameters in the experiments.}
	\label{tab:LGP_param}
\end{table}

\section{Sensor-based feedback optimized with MLC}
\label{Sec:ClosedLoopControl}
 \graphicspath{{./Figures/}}
 \begin{figure}
 	\centering
 	\includegraphics[width=0.45\textwidth,keepaspectratio]{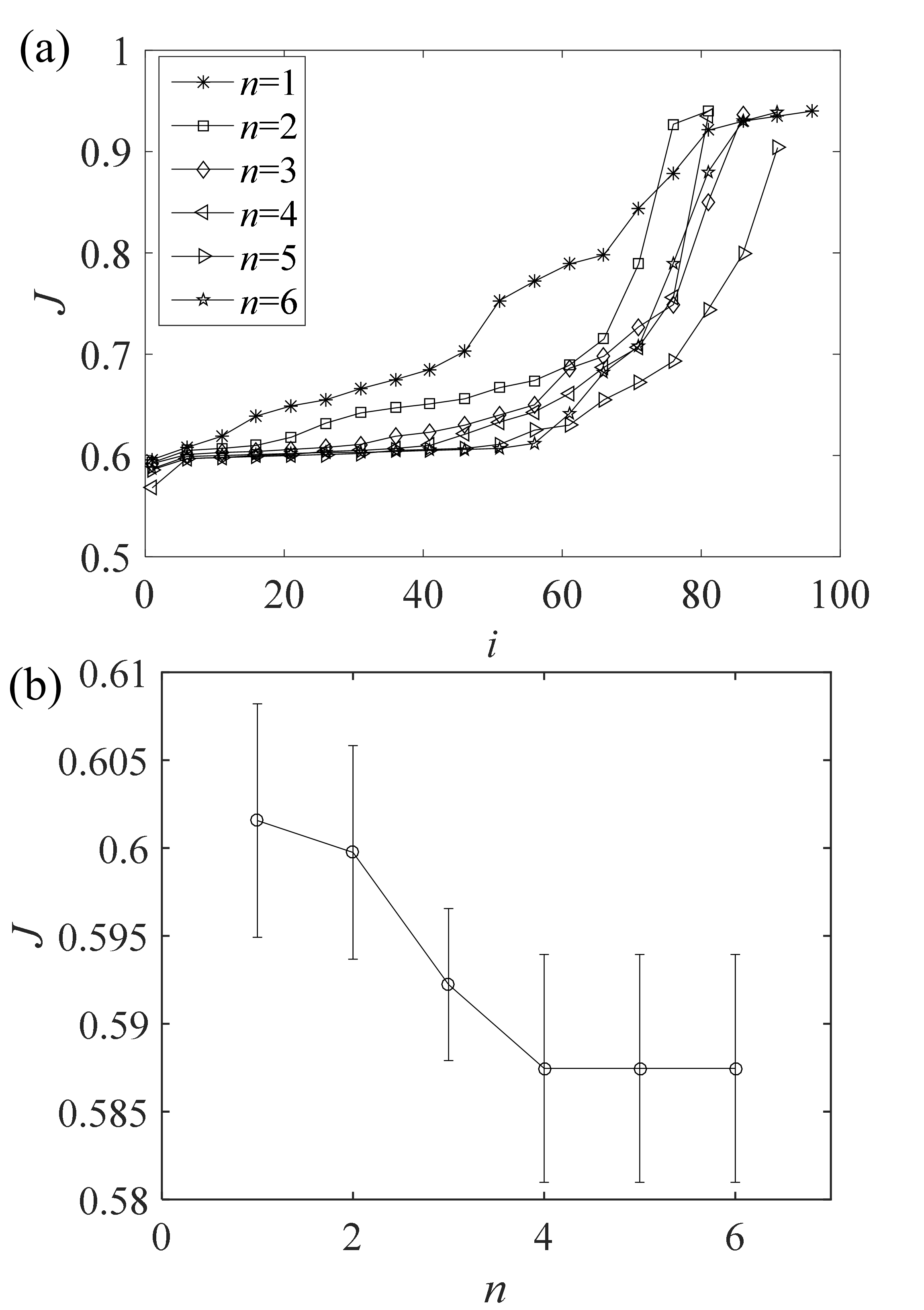}
 	\caption{Learning of the sensor-based feedback with MLC.
        (a) Evolution of the  cost $J$ versus the individual index $i=1,\ldots,100$ for 6 generations $n = 1,\ldots,6$.
        For visual clarity, only every fifth individual is plotted, i.e. $i=1,6,11,\ldots,96$.
        (b) Cost of the optimal individual in the 6 generations $J_1^{(n)}$; $n = 1,\ldots, 6$. The circles correspond to the averaged $J$ values and the error bars mark the standard deviation of repeating evaluations of the optimal control law.}
 	\label{Fig:SensorFeedbackMLC}
 \end{figure}
In this section,
a sensor-based feedback control using MLC \eqref{Eqn:GeneralizedFeedback} is investigated.
Experiments are performed at $Re_D$=8000 and $C_m$=1.2\%.
The $C_m$ of 1.2\% corresponds to the optimum jet mixing performance \citep{Wu2016ef}.
Figure \ref{Fig:SensorFeedbackMLC}a shows the evolution of the cost $J_i^n$
as function of the individual index $i=1,\ldots,100$ for 6 generations ($n=1,\ldots,6$).
For visual clarity, only every $5$th data is displayed, i.e. $i=1,6,11, \dots,96$.
The individuals of each generation
are ordered by $J$ value following \eqref{Eqn:CostFunctionOrder}.
Here, $i = 1$ corresponds to the smallest $J$ and the best control law.
For the first generation $n=1$,
$J$ increases gradually with $i$ starting from the minimum of $J_1^{(1)}\approx0.6$.
With increasing number of generations,
the curve converges to a plateau for the first half of the individuals.
Genetic algorithm breads more and more similar or even identical individuals
with every generation.
At some point,
the learning is converged while the last half of the individuals
explore other new control laws but fail to find better minima.

This convergence is depicted in Figure \ref{Fig:SensorFeedbackMLC}b
showing the performance $J_1^{(n)}$ of the best individual
for each generation $n=1,\ldots,6$.
At $n \geq 4$ the $J$ value is converged  to $0.588$
corresponding to a decay rate of $K=1-J \approx 0.412$.
At $n=6$ the evolution is stopped and the best individual
of the last generation $b_{\rm opt}$ is taken as MLC law:
\begin{equation}
\label{Eqn:SensorFeedbackMLC}
b_{\rm opt}=H (-s_3)
\end{equation}
The actuation fires when the streamwise velocity of the shear-layer sensor at $x/D=3$
is below the average.
Equation \eqref{Eqn:SensorFeedbackMLC} is the simplified version of the equivalent LGP algorithm
$b_{\rm opt}=H \left(0.752 s_3 / (-0.576) \right)$.
The error bar in Fig. \ref{Fig:SensorFeedbackMLC}b displays the standard deviation
of the repeated tests
of the best control law in all the generations.
The error bar significantly decreases with increasing time window.
For MLC, only the approximate relative ordering needs to be preserved
and we save significant measurement time by using a 5 second time window.

 \begin{figure}
 	\centering
 	\includegraphics[width=0.45\textwidth,keepaspectratio]{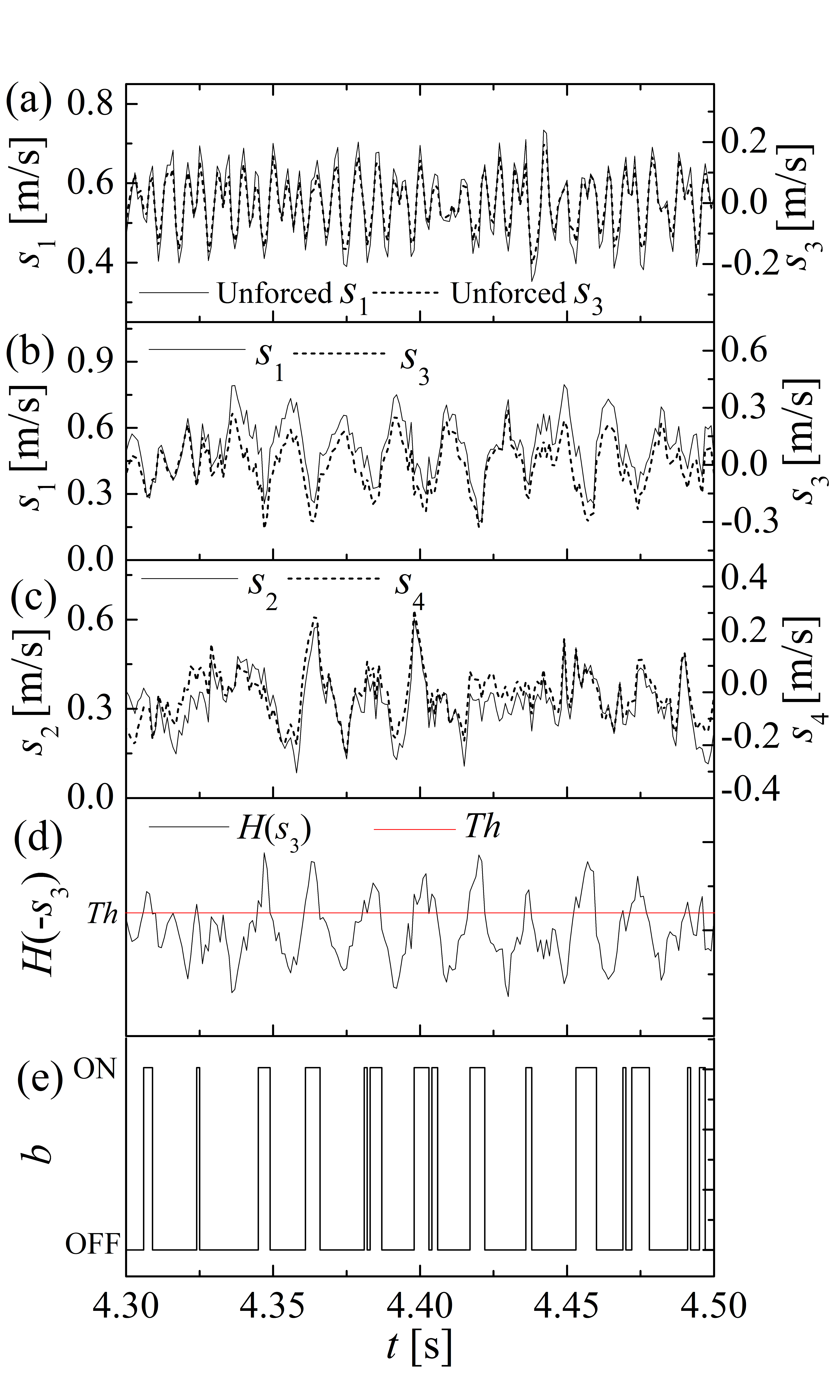}
 	\caption{Sensor-based feedback of MLC---Time history.
         The instantaneous unforced signal $s_1$ and $s_3$ (a), forced signals $s_1$--$s_4$ (b-c), signal of the best control law \eqref{Eqn:SensorFeedbackMLC}(d)
        and the corresponding control signals $b$ (e).}
 	\label{fig:SS}
 \end{figure}
The MLC feedback forcing \eqref{Eqn:SensorFeedbackMLC}
leads to nearly periodic forcing
as displayed in Fig.\ \ref{fig:SS}.
The unforced $s_1$ signal from the shear-layer sensor (Fig. \ref{fig:SS}a)
displays a dominant periodicity at $f_0$ = 135 Hz.
This frequency  corresponds to the coherent shear-layer structures.
The forcing leads to a much  lower frequency of 55 Hz corresponding to a flapping.
This flapping frequency dominates the behavior of the shear-layer sensor at $x/D=3$
and is also visible in the reading of the centerline sensor at $x/D=5$.
Note that $s_3$ follows $s_1$ modulo the short-time averaged mean value.
Ditto for $s_4$ and $s_2$.
The actuation fires for short periods of time
when $s_3$ is negative following \eqref{Eqn:SensorFeedbackMLC}
(see Fig. \ref{fig:SS}b).
In other words, the actuation fires whenever the control law \eqref{Eqn:SensorFeedbackMLC}
is larger than the threshold value $Th$ (see Fig.~\ref{fig:SS}d and e).

 \begin{figure}
 	\centering
 	\includegraphics[width=0.45\textwidth,keepaspectratio]{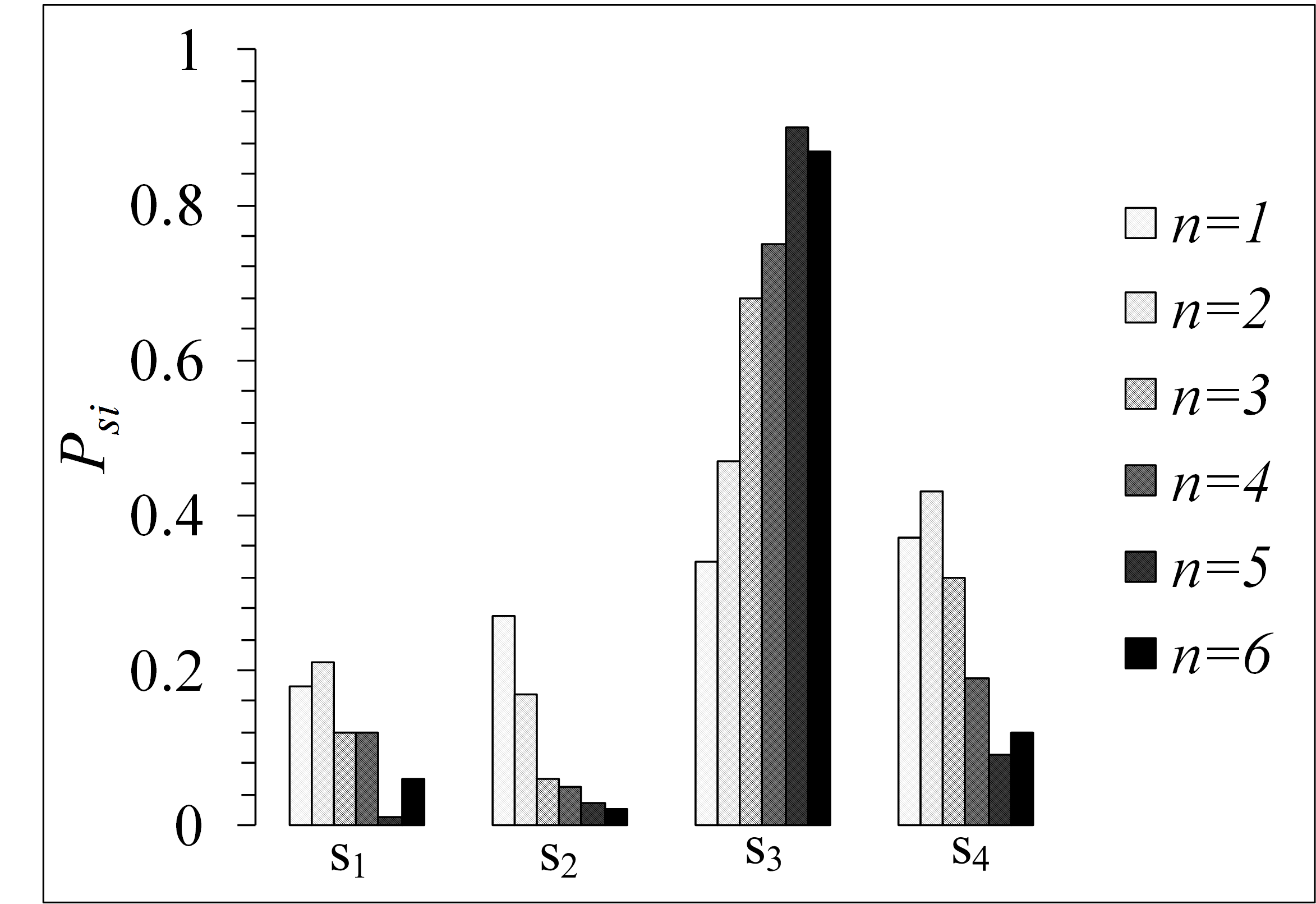}
 	\caption{Sensor-based feedback of MLC---Learning of the sensor selection.
           For each generation $n = 1,\dots,6$, we represent the percentage $P_{si}$ of individuals having $s_i$ in the expression of the $I$ individuals.}
 	\label{fig:Psi}
 \end{figure}
The MLC discovery of $s_3$ as actuation trigger is visualized in Figure \ref{fig:Psi}.
This graph represents the percentage $P_{si}$ of having $s_i$
in the expression of all $I$ control laws for each generation $n$.
With increasing number of generations $n$,
the dominance of $s_3$ becomes more pronounced.
In the last generation, $s_1$, $s_2$ and $s_4$ are neglected in most individuals as  control law input.
Note that multiple appearances of $s_3$ in one individual are counted only once.
This behavior is easily explained by the high efficiency
of periodic forcing for jet mixing via flapping
and by the higher sensitivity of the shear-layer sensor to this frequency.

The convergence of the learning process is displayed in Figure \ref{Fig:SensorFeedbackMLC}.
Following \cite{Kaiser2017ifac},
we also want to learn the `control landscape'
of all tested individuals employing a proximity map.
For that purpose,
we rely on Multi Dimensional Scaling (MDS) \citep{Mardia1979book},
a method classically used to visualize high-dimensional data in a low-dimensional feature space.
Specifically,
we employ classical multidimensional scaling (CMDS)
which is originated from the works of Schoenberg (1935) and Young \& Householder (1938)
and optimally preserves the distances between the data
in the projection from a high-dimensional space
to a low-dimensional feature space.
Our infinite-dimensional objects are  $N=I \times 6 = 600$  control laws.
The control laws are indexed in order of appearance,
i.e.\ $i=1,\ldots,100$ belong to the first generation,
$i=101,\ldots,200$  refer to the second generation, and so on.
Next,  we need to quantify the relative configuration of control laws
with a distance matrix $D=(D_{lm})_{1\leq{l,m}\leq{N}}$.
Here,  $D_{lm}$ denotes the difference between individuals $l$ and $m$.
The square of the distance matrix $\boldsymbol D^2$ is defined by
\begin{equation}%
{D_{lm}^2 = {\left\langle {{{\left| {{b_l}(s) - {b_m}(s)} \right|}^2}} \right\rangle _{l,m}} + \alpha \left| {{J_l} - {J_m}} \right|}
\label{Eqn:Distance}
\end{equation}
The first term represents the difference between the $l$th and $m$th
control laws averaged over the sensor readings
of both actuated dynamics.
Thus, the averaging takes into account the frequency and relevance of the sensor reading.
The second term penalizes the difference of their achieved costs $J$ with coefficient $\alpha$.
This penalization smoothes the control landscape,
i.e. the visualization of the cost $J$ in the feature space.
The penalization parameter $\alpha$ is chosen
so that the maximum variation of the first
and second term of \eqref{Eqn:Distance} are equal.
Thus, the dissimilarities between control laws and between the cost functions
have comparable weights in the distance matrix $D_{lm}$.
For further details, please refer to \cite{Duriez2016book}.
The aim of CMDS is to find a centred representation
of points $\Gamma  = \left[ \vec{\gamma_1},\vec{\gamma_2},\dots,\vec{\gamma_N} \right ]$
with $\vec{\gamma_1}$, $\vec{\gamma_2}$,\dots, $\vec{\gamma_N}$ $\in$ $\mathbb{R}^2$,
such that the pairwise distances of the feature points
are---in a well-defined sense---optimally close to the original distances,
i.e.\ $\left\|\vec{\gamma}_l-\vec{\gamma}_m\right\|_2\approx D_{lm}$.

 \begin{figure}
 	\centering
 	\includegraphics[width=0.45\textwidth,keepaspectratio]{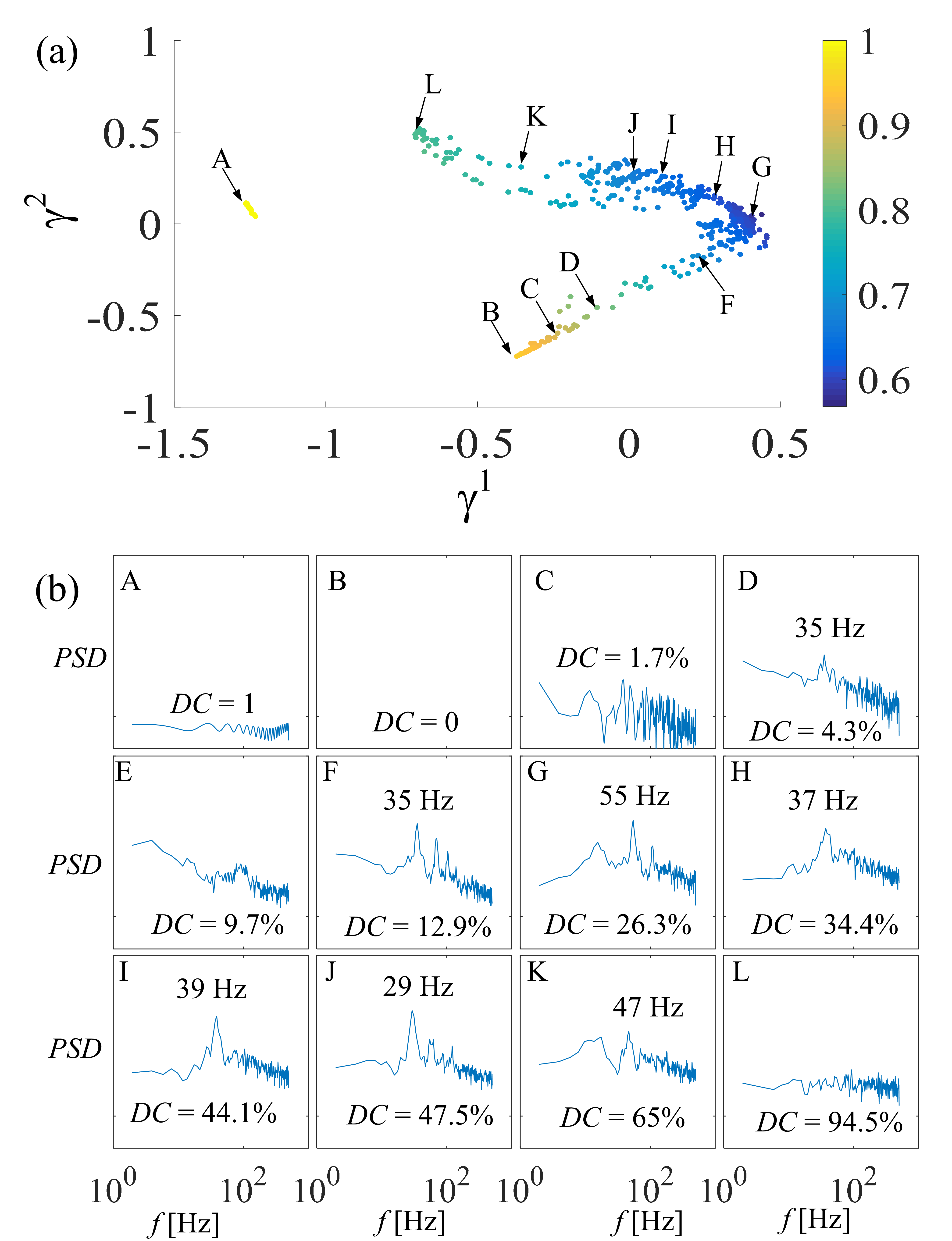}
 	\caption{Proximity map of all tested control laws for sensor-based feedback with MLC.
(a) Visualization of (dis)similarity associated with the entire collection (600 individuals) of the sensor-based feedback control laws.
Each dot represents an individual control law and the distance between two control laws approximates their respective dissimilarity.
The color scheme corresponds to the percentile rank of the control laws with respect to their performance $J$.
The best performing individuals is indicated by $G$.
 (b) the corresponding power spectrum and duty cycles of control laws $A-L$.}
 	\label{Fig:ProximityMapSensorFeedback}
 \end{figure}
Figure \ref{Fig:ProximityMapSensorFeedback} shows the proximity map of all control laws $(N=600)$
in a two dimensional plane $\left ( \gamma^1, \gamma^2 \right) \in \mathbb{R}^2$
and the power spectral density of thee selected control laws $A$-$L$, respectively.
The $\gamma^1$ and $\gamma^2$ represent the coordinates in the two dimensional plane.
Each dot represents one control law and the corresponding $J$ value is color-coded.
The optimal law $b_{\rm opt}$ is indexed by $G$.
All individuals are close to a $V$-shaped curve
except continuous blowing with  duty cycle $DC=1$ as isolated point $A$.
On the curve,
the $DC$ increases constantly from 0 to 94.5\% for control laws $B$ to $L$.
The second feature coordinate $\gamma^2$ clearly correlates
with the duty cycle ($DC$) of the control signal.
The first feature coordinate $\gamma^1$ appears to correlate with spectral characteristics.
The maximum frequency of the best performing law $G$ is on the rightmost side of the curve
while the other individuals have lower dominant frequencies.
The proximity map reveals that MLC explores multi-frequency actuation mechanisms
and arrives at a dominant periodicity with low duty cycle of 26.3\%
and new flapping frequency of $55$ Hz.
This frequency of $55$ Hz does not equal the optimal frequency (i.e.~67 Hz) achieved using periodic forcing method \citep{Fan2017aiaaj}. One possible reason is the influence of turbulence, which makes the the optimal frequency and duty cycle hardly be searched by sensor-based feedback control law \eqref{Eqn:SensorFeedbackMLC}. Detailed discussion will be given in Sect.~\ref{Sec:Discussion:Feedback}.

\section{Multi-frequency forcing optimized with MLC}

\label{Sec:OpenLoopForcing}
\begin{table}
	\centering	
	\begin{tabular}{l|lllllllll}
		\hline
$h_i$ & $h_1$ & $h_2$ & $h_3$ & $h_4$ &$h_5$ & $h_6$ & $h_7$ & $h_8$ & $h_9$ \\
$f_i$ [Hz] & 20 & 50 & 59 & 67 & 77 & 91 & 111 & 143 & 200 \\
         \hline	
	\end{tabular}
	\caption{Frequencies of the harmonic functions $h_i(t) = \sin(2\pi f_it)$
                 used as inputs for the MLC law yielding multi-frequency actuation.}
	\label{tab:hi}
\end{table}
The sensor-based feedback control results distill
nearly periodic forcing as best actuation.
However, the possibility of feedback control to give rise to strict periodic forcing
is mitigated by the low-frequency drifts and high-frequency noise.
Moreover, in some MLC studies,
open-loop multi-frequency forcing
has been shown to outperform both periodic forcing
and sensor-based feedback \citep{Li2017ef}.
This motivates the use of MLC to optimize open-loop multi-frequency actuation.
Here, a range of harmonic functions are used as inputs of control laws.
Table \ref{tab:hi} provides the chosen harmonic functions $h_i(t)$.

 \begin{figure}%
 	\centering
 	\includegraphics[width=0.45\textwidth,keepaspectratio]{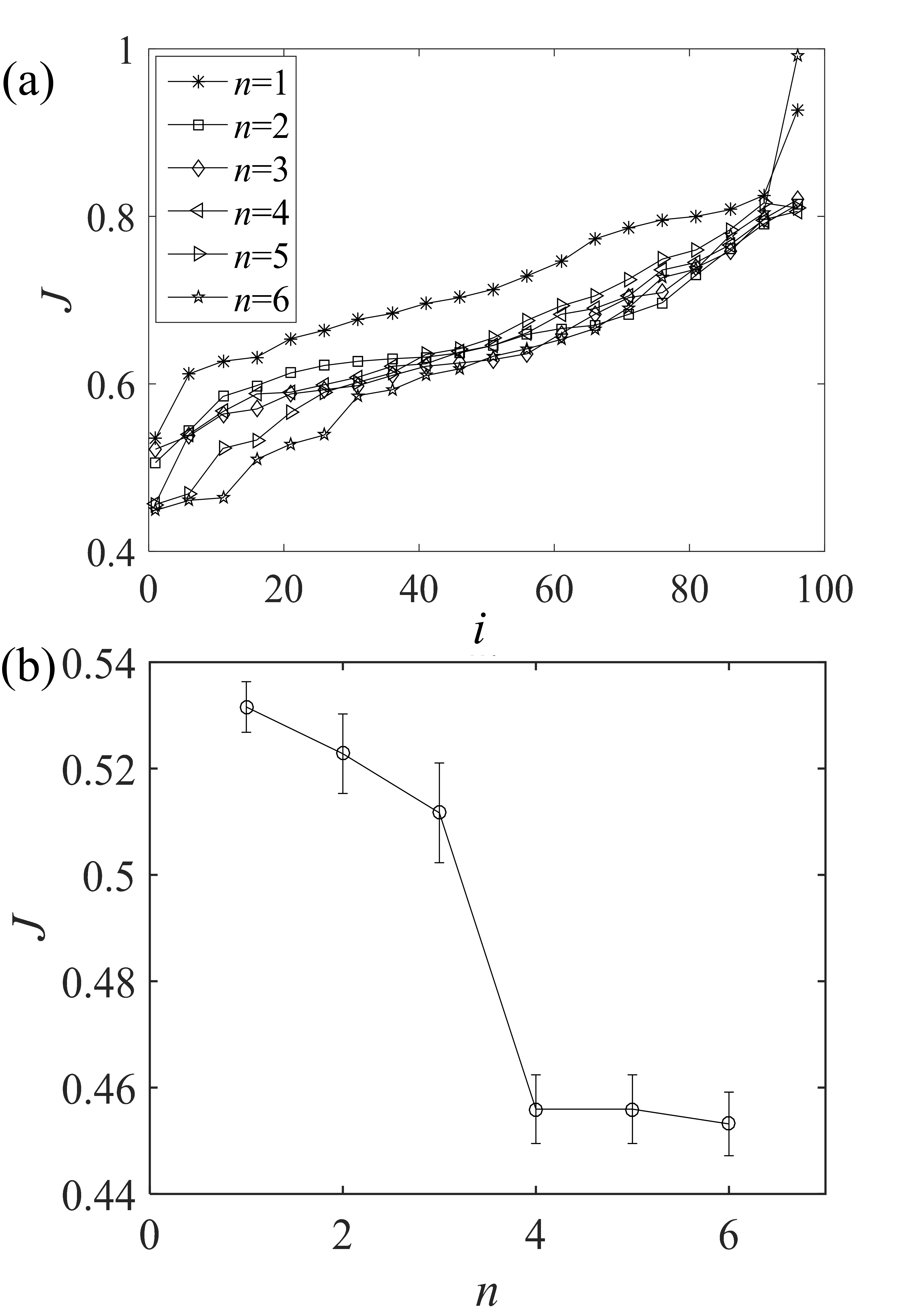}
 	\caption{Learning multi-frequency actuation with MLC.
 	The cost function $J_i^{(n)}$ is displayed like in figure \ref{Fig:SensorFeedbackMLC}.}
 	\label{Fig:MultiFrequencyMLC}
 \end{figure}
Figure \ref{Fig:MultiFrequencyMLC}a illustrates the evolution of
the cost $J$ as function of the individual index $i$ for 6 generations.
As for the sensor-based feedback,
we plot only every $5$th individual for visual clarity.
Interestingly, the increase of $J$ with $i$ is more steep than that for the sensor-based feedback,
indicating that the minimum is less populated.
One reason may be that the number of control law arguments has more than doubled
and the search space is in some vague sense `larger'.
Another reason is that the harmonic functions are by construction
less correlated than the sensor signals.
This reduced correlation has frequently been found
to be associated with the increase of learning time.
A particularly noteworthy characteristic of this MLC run is the sudden jump
of the best cost value $J_1^{(n)}$ from $0.510$ to the converged value of $J_1^{(4)}= 0.458$ in the fourth generation (Fig.\ \ref{Fig:MultiFrequencyMLC}b), indicating that a new minimum may have been found from generation 3 to 4.
This value is 24\% better than 0.588 of the sensor-based feedback,
and corresponds to a decay rate of $\overline{K}=0.542$.
The error bars of open- and closed-loop MLC are similar to each other.

 \begin{figure}%
 	\centering
 	\includegraphics[width=0.45\textwidth,keepaspectratio]{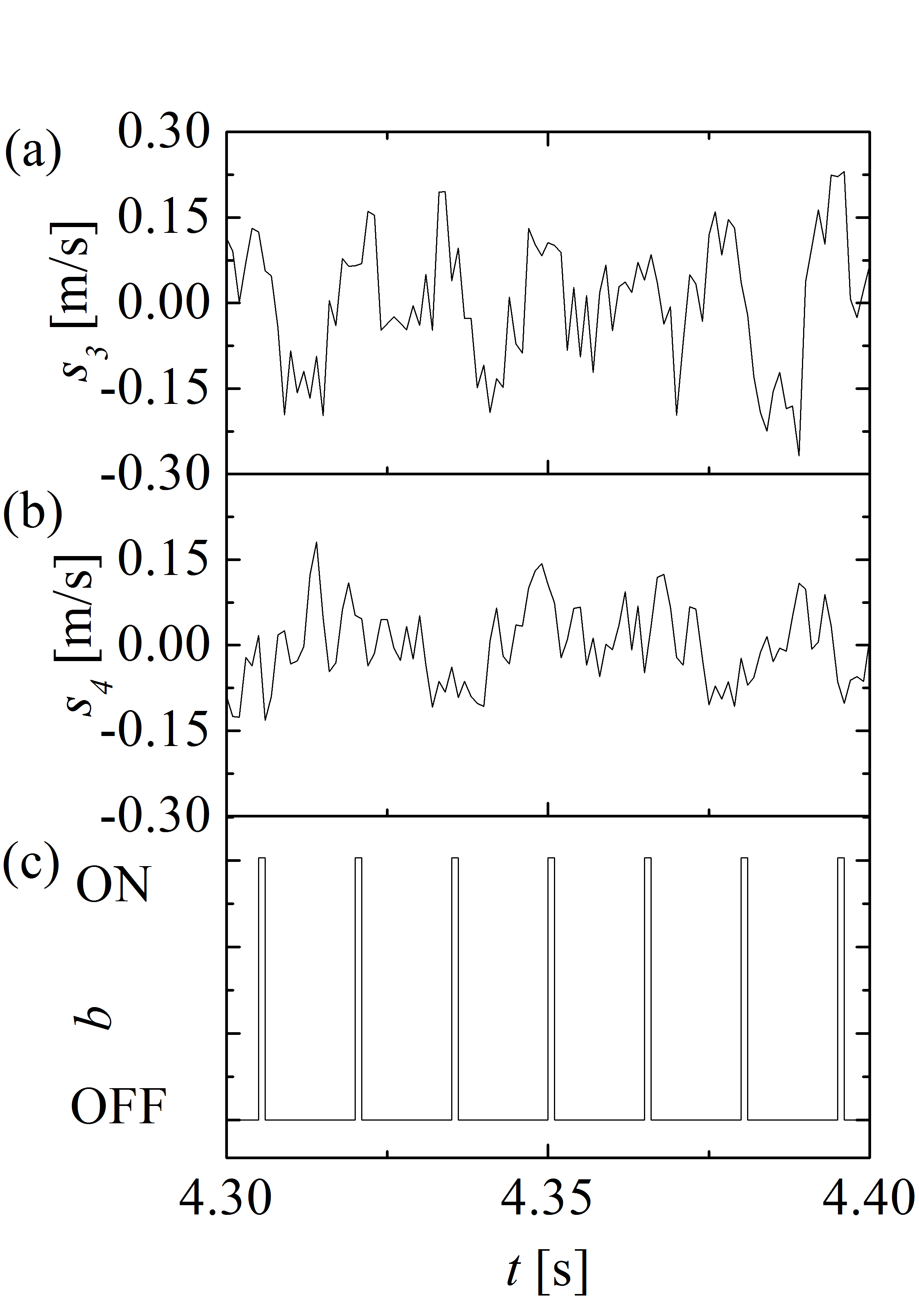}
 	\caption{Multi-frequency forcing with MLC---Time histories.
                 The instantaneous sensor signals $s_3$ (a), $s_4$ (b)
                 and control signals $b$ (c) are displayed for the best control law \eqref{Eqn:OpenLoopMLC}.}
 	\label{Fig:MultiFrequencyDynamics}
 \end{figure}
The resulting MLC law reads
\begin{equation}
\label{Eqn:OpenLoopMLC}
b_{\rm opt} = H \left( \log(\left|((-0.646-\log(\left|h_4\right|)^2))\right| \right)
\end{equation}
and corresponds to periodic forcing with a frequency of $f_e= 67$ Hz and a duty cycle of 7\%.
 Figure \ref{Fig:MultiFrequencyDynamics} illustrates the corresponding actuation command
and the resulting sensor fluctuations in the shear layer $(s_3)$ and on the centerline $(s_4)$
 in a few period interval.
The actuation is strongly correlated to low $s_3$ values,
that is, the mechanisms of MLC for open- and closed loops are similar.
The $s_3$ signal tends to follow the actuation with a minor delay.
The sensor  $s_4$ associated with the cost function shows little response to actuation.

 \begin{figure}%
 	\centering
 	\includegraphics[width=0.45\textwidth,keepaspectratio]{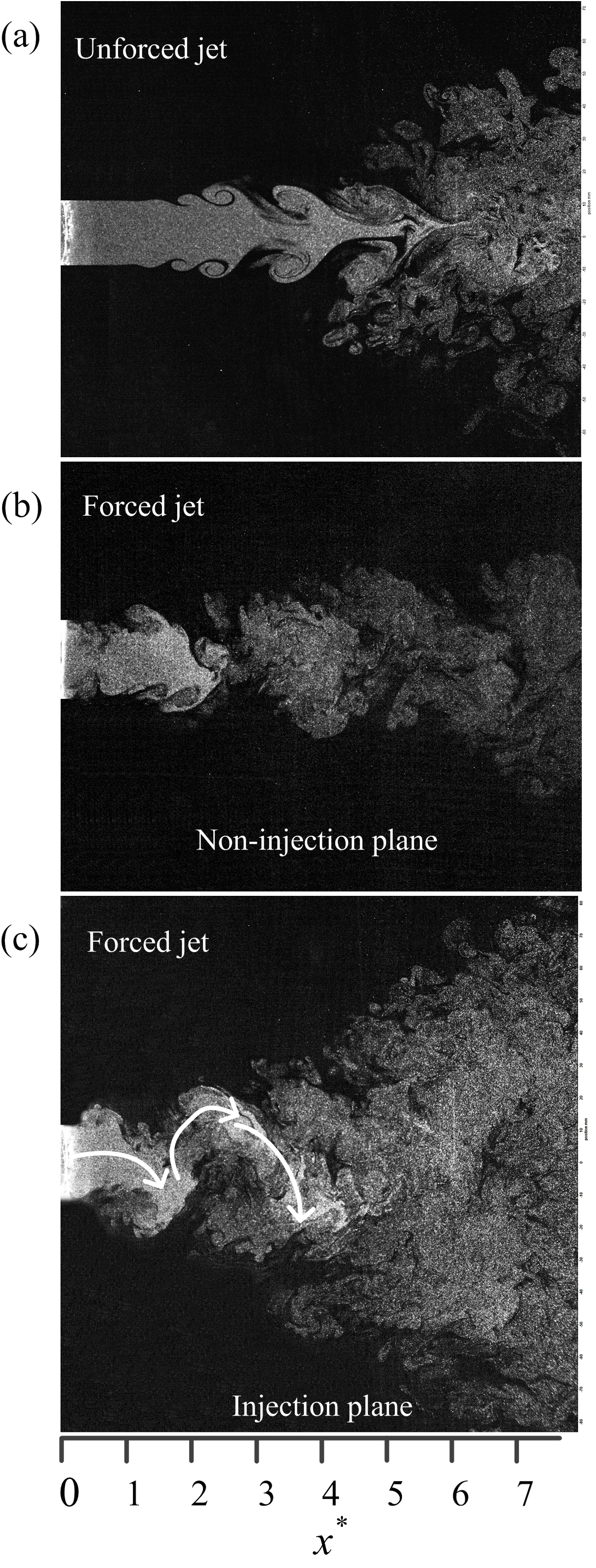}
 	\caption{Photographs from flow visualization of the unforced benchmark and the optimal actuation.
                 Flow is from the left to right.
                 Comparison in the typical flow structure between the uncontrolled jet (a)
                 and the controlled jet $(C_m = 1.2 \%, f_e = 67 {\rm Hz}, DC=7\%)$ in the non-injection (b) and injection planes (c). }
 	\label{fig:s5 FV}
 \end{figure}
Figure \ref{fig:s5 FV} shows typical photographs
in the injection $(x, z)$
and non-injection $(x, y)$ planes. The unforced jet ( Fig.~\ref{fig:s5 FV}a) is, by symmetry, similar in both planes.
The smoke clearly distills high-frequency shear-layer vortices
which are consistent with the observed $135$ Hz shear-layer signal of Fig.~\ref{fig:SS}a.
Further visualizations in the $(y,z)$ plane (not shown here)
are consistent with axisymmetric ring vortices.
Figures \ref{fig:s5 FV} b and c show the flow visualization data under the optimal periodic forcing \eqref{Eqn:OpenLoopMLC}.
The photograph in the injection plane displays a strong flapping motion
which leads to large dispersion of smoke in the transverse direction.
That in the non-injection plane indicates no increase in the transversal mixing.
Yet, 
the photographs in both planes indicate that the length of the potential core is about 2D.
This value is to be compared with the continuous smoke
on the centerline until at least 4 diameters for the unforced jet.
These observations are consistent with the observed decrease of the chosen cost function.
This flapping, characterized by greatly enhanced entrainment in the non-injection plane and very rapid spread in the injection plane, has also been observed in \cite{Yang2016pmpes}.
In their study, the jet manipulation was performed with two asymmetrically arranged unsteady minijets.
All results convincingly demonstrate
that the flapping motion is responsible for the rapid decay
of the centreline mean velocity (Fig. \ref{fig:s5 FV}).

 \begin{figure}%
 	\centering
 	\includegraphics[width=0.45\textwidth,keepaspectratio]{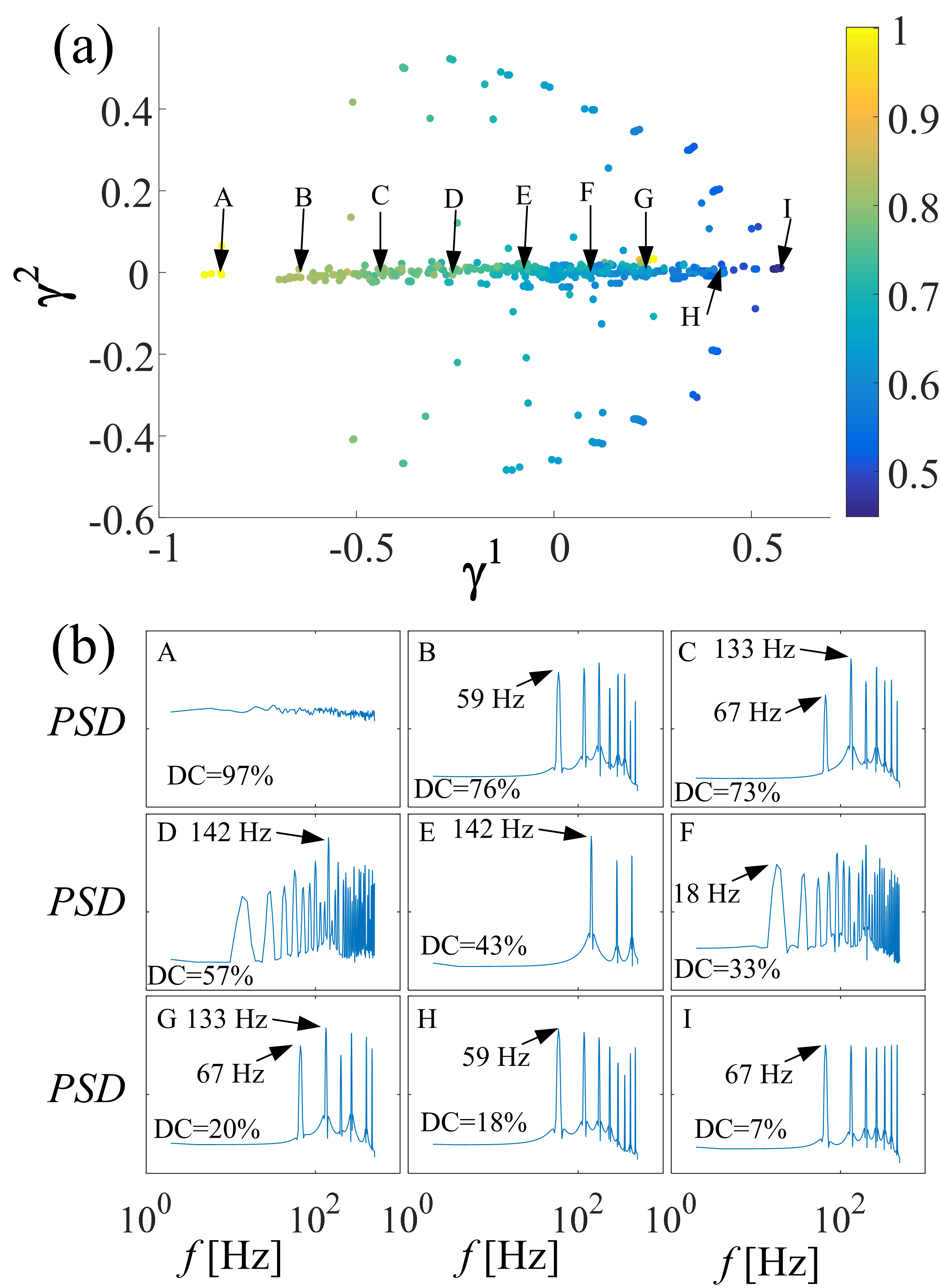}
 	\caption{Proximity map of MLC with multi-frequency forcing
in analogy to Fig.~\ref{Fig:ProximityMapSensorFeedback}. For details see text.}
 	\label{Fig:ProximityMapMultiFrequency}
 \end{figure}
Similarly to Sect.~4,
the control landscape is given in Fig. \ref{Fig:ProximityMapMultiFrequency}a.
Figure \ref{Fig:ProximityMapMultiFrequency}b shows the power spectral density functions for selected control laws $A$--$I$.
The 
coordinate $\gamma^1$
is strongly correlated to $DC$.
From $A$ to $I$,
the $DC$ varies from 97\% (left) to 7\% (right),
while the cost $J$ decreases from 1 to 0.458. The optimal law $b_{\rm opt}$ is indexed by $I$.
These results indicate again that  $DC$ plays an important role in control performance.


\section{Discussion of open versus closed-loop control}
\label{Sec:Discussion}

Sections \ref{Sec:ClosedLoopControl}
and \ref{Sec:OpenLoopForcing}
reveal arguably surprising features of machine learning control
and of the optimal actuation mechanism.
In the following, we present three aspects:
the learning rate for periodic forcing
(Sect.~\ref{Sec:Discussion:Learning}),
the poor performance of feedback control
(Sect.~\ref{Sec:Discussion:Feedback}) and
the convergence against pure periodic forcing
without other frequency components
(Sect.~\ref{Sec:Discussion:MultiFrequencyForcing}).

\subsection{Learning rate of MLC for optimal periodic forcing}
\label{Sec:Discussion:Learning}
In Sect.~\ref{Sec:OpenLoopForcing},
MLC performs a global search over many frequencies and many duty cycles
and arrives at pure periodic forcing after testing only 400 individuals.
From this and the previous study \citep{Wu2016ef},
the characterizing frequency and duty cycle of this periodic forcing
could not be improved further by a systematic search.

 \begin{figure}%
 	\centering
 	\includegraphics[width=0.45\textwidth,keepaspectratio]{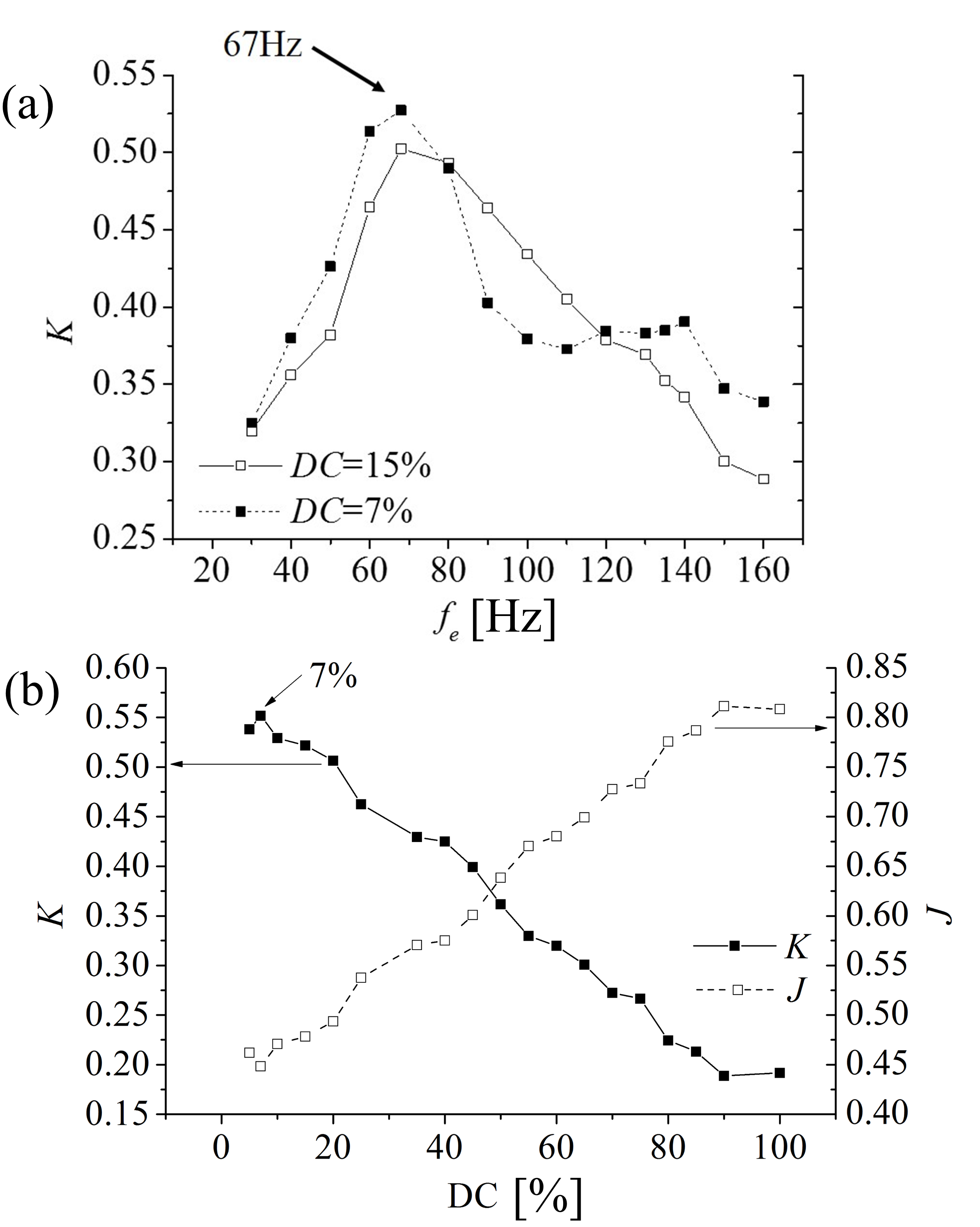}
 	\caption{Performance of periodic forcing.
              (a) The dependence of $K$ on $f_e$ at $Re_D$ = 8000, $C_m \approx 1.2\%$. (b) The dependence of $K$ on $DC$ at $Re_D$ = 8000, $C_m = 1.2\%$ and $f_e$ = 67 Hz.}
 	\label{Fig:PeriodicForcingParametrics}
 \end{figure}
 \begin{figure}%
 	\centering
 	\includegraphics[width=0.45\textwidth,keepaspectratio]{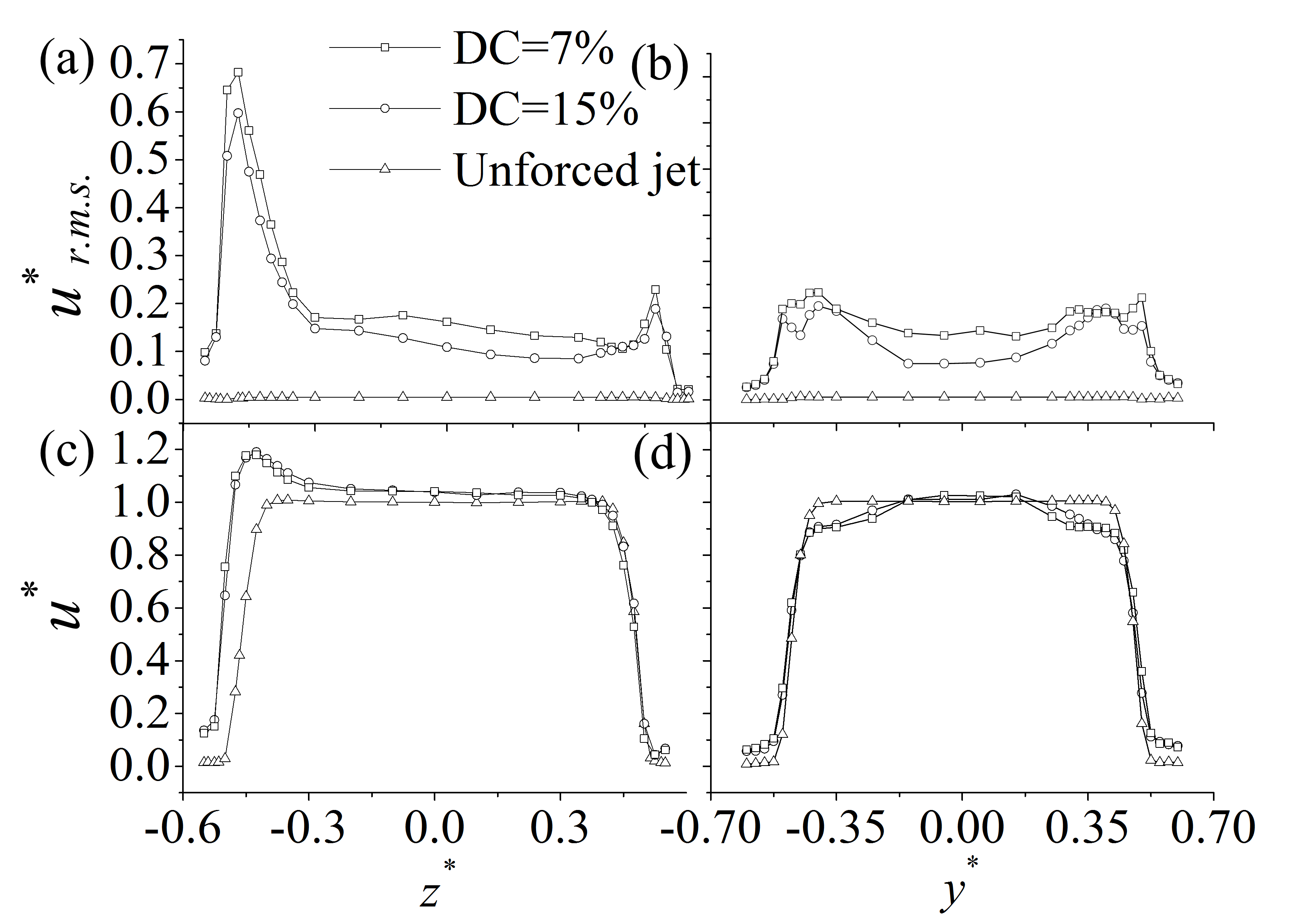}
 	\caption{Optimal periodic forcing: Radial distribution of (a, b) turbulent intensity ${u^*}_{\rm r.m.s.} = u_{\rm r.m.s.}/U_j$ and (c, d) $u^*= u/U_j$ at $x^*=0.05$ of the unforced jet and manipulated jet at $DC = 7\%$ and $DC = 15\%$ in $x-z$ and $x-y$ plane, $C_m = 1.2\%$ and $f_e$ = 67 Hz.}
 	\label{Fig:PeriodicForcingOptimal}
 \end{figure}
Moreover, the decay rate $K$ is 8\% larger as compared to the previously achieved mixing,
where $DC$ was fixed at 15\% and $C_m$ and $f_e$
were optimized using a dual-input/single-output extremum-seeking controller \citep{Wu2018aiaaj},
yielding $C_m = 1.2\%$ and $f_e$=67 Hz (see Fig.\ref{Fig:PeriodicForcingParametrics}a, $DC$=15\%).
The MLC-optimized open-loop actuation is strictly periodic.
The control law \eqref{Eqn:OpenLoopMLC}
contains only the excitation frequency $h_4$.
This frequency is the same as $f_e=67$ Hz as identified by \cite{Fan2017aiaaj}. Note that $f_e$ = $f_0/2$,
i.e.\ half the unforced shear-layer frequency.
The improved control performance is predominantly caused by duty cycle $DC$
which is reduced to $7$\% when MLC is employed.
The optimality of $DC$ at fixed exication frequency
is demonstrated in Fig.~\ref{Fig:PeriodicForcingParametrics}b.
Figure \ref{Fig:PeriodicForcingOptimal} evidences that, with a fixed $C_m$, a lower $DC$ yields larger changes of the main jet, especially for the standard deviation $u_{\rm r.m.s.}$.
Note that the employed electromagnetic-valves do not open when the $DC$ is less than 5\%.

This performance can be considered quite impressive
for an evolutionary learning algorithm.
These algorithms are powerful for exploration
but perform less well for exploitation,
the realm of gradient-based parameter optimization.
Moreover, 400 individuals to convergence
may be compared with a systematic parameter variation
over two parameters.
Let us assume 400 cost functions are 
evaluated with a systematic equidistant sampling
using 20 frequencies in the range from 0 to $200$ Hz
and 20 duty cycles from zero to unity.
The closest result $f_e$=60 Hz and $DC$=10\%
would be far away from the optimal parameters and
yield a significantly worse performance.

MLC may also be compared with local gradient search---at the expense of potentially arriving in a suboptimal minimum.
A two-parameter extremum seeking
and a two-parameter simplex search can be expected
to require $O(100)$ test runs worth of measurement time.
Summarizing, MLC is a competitive two-parameter optimizer,
a task which is not a typical application of an evolutionary learning algorithm.
In addition, MLC has operated in much larger search space
of multi-frequency laws.
A side benefit of this search is
that the performance advantages of sensor-based feedback
and of non-periodic actuation have been assessed.

\subsection{No performance benefits of sensor-based feedback}
\label{Sec:Discussion:Feedback}
\begin{figure}
\centering
\includegraphics[width=0.45 \textwidth]{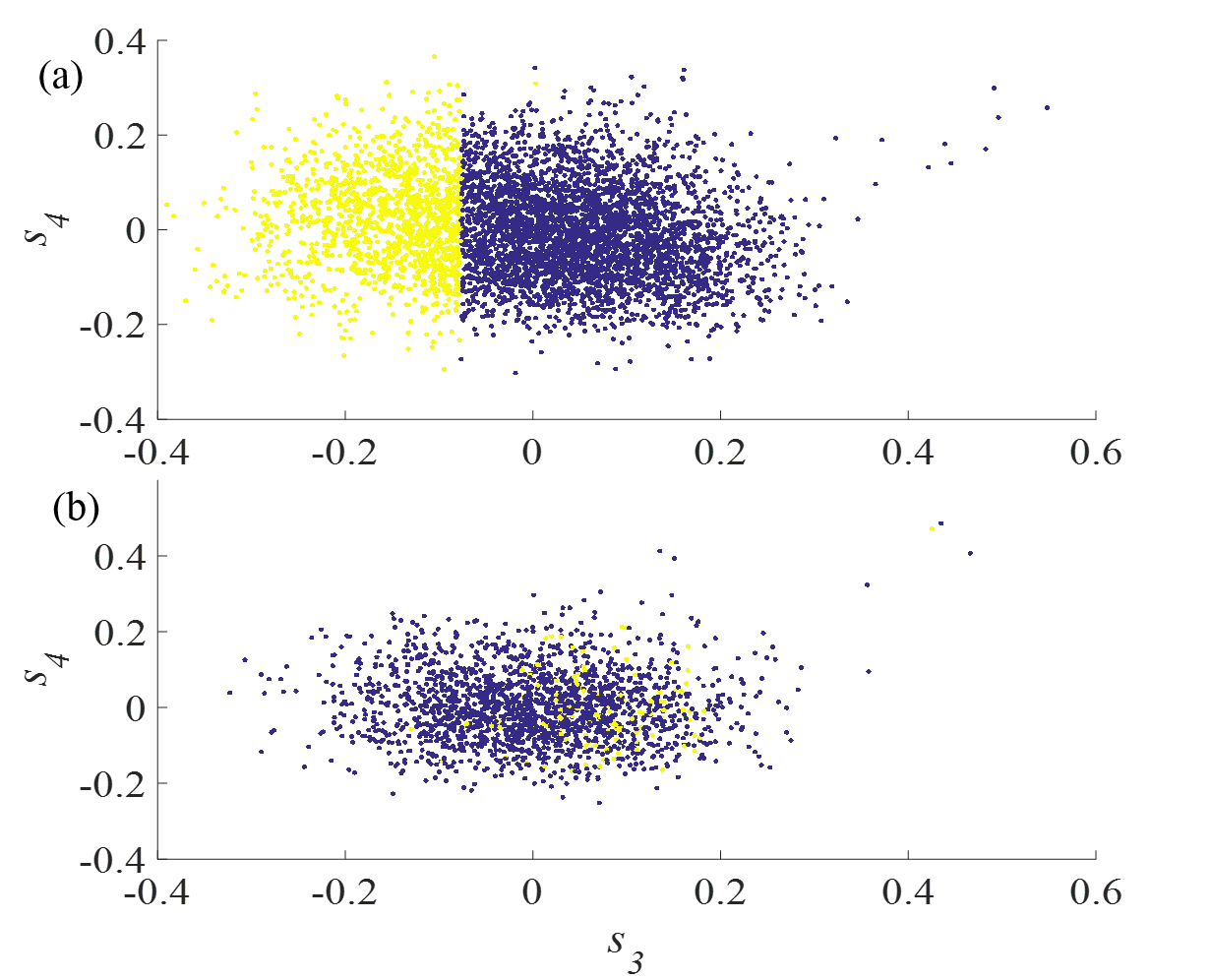}
\caption{Analysis of the optimal MLC-based open-loop (a) and closed-loop control (b).
Both figures show measured sensor signals $s_3$ and $s_4$
and indicate the forcing with solid yellow bullets for actuation and blue bullets for the off-phase.}
\label{Fig:SensingActuation}
\end{figure}

The performance of sensor-based feedback is surprisingly low.
In the following,
we explore if sensor-based feedback $b=B(s_1,s_2,s_3,s_4)$ could have mimicked
the best open-loop control.
The MLC feedback law $b=B_{\rm opt}(s_3)$ employes only $s_3$, 
i.e. 
the velocity fluctuation at $x=3D$.
Based on Fig. \ref{fig:Psi}, we also include the fluctuation at $x=5D$
in the analyzed control law $b=B(s_3,s_4)$,
ignoring $s_1$ and $s_2$ on grounds
that they are hardly used in later MLC generations.
Figure \ref{Fig:SensingActuation} shows the actuation $b$
at measured sensor signals $s_3$ and $s_4$ both,
for MLC-based open- and closed-loop actuation.
Figure \ref{Fig:SensingActuation}a shows clearly the $s_3$-dependent `firing' criteria.
However, in Fig.~\ref{Fig:SensingActuation}b, the optimal periodic forcing
seems hardly to be correlated with $s_3$ and $s_4$.
The  ON and OFF actuation phases
appear statistically distributed over the whole data region.
This mixing of two actuation states
strongly indicates that the optimal periodic forcing
cannot 
be found by the ansatz $b=B(s_3,s_4)$.

One should also keep in mind
that the optimal duty cycle of $7$\%
is a very small `firing' window.
Feedback may miss this window due to the high-frequency noise and low-frequency drifts.
Any `miss' of this window implies a performance reduction.
Thus, the lower performance of sensor-based feedback is---in hindsight---not surprising for such short duty cycles.

The performance has been improved by moving the second hot-wire closer to the nozzle exit.
Further performance increases have been explored
with Morlet-filtering \citep{Bourgeois2013jfm}
or other forms of signal processing.
In all these cases, we did not find a sensor-based feedback
which has outperformed periodic forcing or was similar.

The ON/OFF actuation with small duty cycles
makes the presented optimal jet mixing different
from other experimental turbulence control studies
where in-time phasor control has outperformed working periodic forcing \citep{Pastoor2008jfm}.
Moreover, the ON/OFF actuation leaves little room for a slow parameter adaption.
However, there were also cases in which the optimal high-frequency forcing
has outperformed MLC-based sensor-based feedback \citep{Parezanovic2016jfm,Li2017ef}.
This is not surprising as the signal-to-noise ratio of  high-frequency actuation effects
versus broadband frequency distribution tends to be very poor.
In \cite{Li2017ef} only one of 16 pressure sensors
was able to resolve the actuation frequency and its phase,
but only with a two period delay.

Following \cite{Li2017ef},
we have also explored generalized control laws $b=K(s_1,\ldots,s_4,h_1,\ldots,h_9)$,
i.e.\ Eq.~\eqref{Eqn:GeneralizedFeedback}.
comprising sensor-based feedback and multi-frequency forcing.
This ansatz has increased the convergence time of MLC
and also yielded the same periodic forcing.

\subsection{No performance benefits from multi-frequency forcing}
\label{Sec:Discussion:MultiFrequencyForcing}

Literature contains many experimental flow control studies
in which multi-frequency forcing has outperformed periodic forcing
at the same or similar amplitude.
In \cite{Li2017ef}, MLC has identified a two-frequency forcing
which has improved the drag reduction of a car model
as compared to optimized periodic forcing.
In \cite{Chovet2017ifac},
a similar observation was made for the mixing enhancement
behind a backward-facing step.
Numerous reports show how sub- and subsubharmonic forcing components
increase the mixing layer width by triggering more early vortex pairing \citep{Monkewitz1988jfm,Coats1997pecs}.
Hence, the performance of pure periodic forcing
for optimal jet mixing is initially surprising.

However, forcing augments jet mixing
by inducing a strong asymmetric flapping of the jet
in the plane with the minijet actuator.
This is a comparably simple mechanism,
like the excitation of pendulum motion with a kick near the lower equilibrium point.
It is difficult to perceive how
scheduling the `firing' non-periodically should improve such an inherently periodic  phenomena.
The mixing layer, for instance, is much more complex
by incorporating multiple vortex merging
and three-dimensional structures,
making multiple frequencies more advantageous.

\section{Conclusions}
\label{Sec:Conclusions}
In this experimental study,
we maximize jet mixing using one minijet actuator
and two hotwire sensors---advancing past closed-loop control studies by the group \citep{Wu2016ef}.
The control law ansatz comprises a multi-frequency forcing,
sensor-based feedback and combinations thereof following \cite{Li2017ef}.
From this large search space,
machine learning control (MLC) has identified periodic forcing
with short duty cycle as optimal.
The mixing is quantified by the averaged streamwise velocity decay rate
at five diameters downstream on the symmetry axis.
The achieved mixing is  better than in a previous study by \cite{Wu2016ef}
with extremum seeking control since a better (smaller) duty cycle was found.
MLC performed optimization in only 4 generations with 100 control laws in each,
i.e.\ 400 runs with 5 seconds evaluation for each run.
The frequency and duty cycle identified by MLC
could not be improved further with parametric studies.
Summarizing,
the learning time is comparable to alternative optimization of periodic forcing,
e.g.\ testing 50 different frequencies for 20 different duty cycles.

In addition to identifying the optimal periodic forcing,
MLC indicates that neither additional forcing frequencies
nor employing sensor-based feedback improves mixing further.
Both implications may initially be surprising
but can---in hindsight---easily be explained.
Numerous turbulence control experiments show how multi-frequency forcing outperforms periodic forcing.
In \cite{Li2017ef}, MLC identified
that multi-frequency forcing is more effective for drag reduction
of a car model than the optimized periodic forcing.
MLC also found multi-frequency forcing to outperform optimized periodic actuation
in the reduction of a recirculation zone behind a backward facing step \citep{Chovet2017ifac}.
The list of similar observations can easily be extended \citep{Coats1997pecs}. 
Multi-frequency forcing is a very large superset of periodic forcing.
In case of  jet mixing,
the underlying  mechanism is a flapping in the plane containing the minijet.
It seems that this simple mechanism cannot be improved by other frequencies.
Moreover, a single actuator may not trigger other mechanisms
which may be based on other different frequencies.

The poor performance of sensor-based feedback
for an oscillatory mechanism is also initially surprising
in light of a common experience of turbulence control experiments:
If periodic forcing  improves a performance, feedback can generally improve it further.
This feedback may adjust in-time the phase of actuation
to flow events \citep{Pastoor2008jfm}
or may perform a slow adaption of a forcing parameter.
However,
in case of the jet mixing neither phasor control nor parameter adaption can be expected to work.
The mixing is critically depending on a short `firing' time in a narrow time interval.
Any feedback `firing' policy will be mitigated by the low-frequency drifts and by high-frequency noise.
In addition, the boolean on-off nature of control
excludes the possibility of an amplitude adaptation.
In hindsight, the poor performance of sensor-based feedback in comparison with the optimal periodic forcing
can be expected if very short duty cycles are necessary for good actuation performance.
We did improve sensor-based feedback by several measures,
e.g.\ optimization of the sensor position or other sensor filters.
However, in none of these experiments the performance of periodic forcing has been reached.

The current jet mixing study reveals that a simple periodic actuation appears to be best.
MLC simultaneously optimizes the forcing parameters at a highly competitive learning rate
and seems to exclude performance increases by feedback and new frequencies.
Subsequent studies of the authors concern jet mixing enhancement in the same facility,
but employing all six minijets.
In this case, MLC is found to yield a dramatic performance increase
with a non-harmonic and non-symmetric forcing not reported in literature so far.
\hbox{MLC} has a large untapped potential in turbulence control applications.
We actively pursue a  MLC generalization which learns not only the optimal control law
but also the corresponding control-oriented model within few hundred or few thousand short test runs.

\begin{acknowledgements}
This work is supported
by a public grant overseen by the French National Research Agency (ANR)
as part of the ``Investissement d’Avenir'' program, through the  ``iCODE Institute project'' funded by the IDEX Paris-Saclay, ANR-11-IDEX-0003-02,
by the ANR grant 'ACTIV\_ROAD'.
The thesis of RL is supported by the OpenLab
Fluidics between PSA Peugeot-Citro\"en and Institute Pprime
(Fluidics@poitiers). The financial support of NSFC via grant (approval no. 91752109) is acknowledged.

We appreciate valuable stimulating discussions with
Steven Brunton, Camila Chovet,  Eurika Kaiser, Laurent Keirsbulck, Nathan Kutz, Richard Semaan and
the French-German-Canadian-American pinball team:
Guy Yoslan Cornejo-Maceda, Nan Deng,  Fran\c{c}ois Lusseyran, Robert Martinuzzi, Cedric Raibaudo and Luc Pastur.

\end{acknowledgements}


\bibliographystyle{spbasic}
\bibliography{Main_Bernd,Main_Wu}

\end{document}